\documentclass[prc,twocolumn,nofootinbib,superscriptaddress]{revtex4-1}

\usepackage{graphicx,amsmath,amssymb,bm,tabularx}
\usepackage{multirow,dcolumn}

\usepackage{xcolor}

\newcommand {\Eq} [1] {Eq.~(\ref{#1})}

\newcolumntype{.}{D{.}{.}{2.3}}

\begin{document}

\title{$\alpha$-nucleus optical potentials from chiral effective field theory $NN$ interactions}

\author{V.\ Durant}
\email[Email:~]{vdurant@uni-mainz.de}
\affiliation{Institut f\"ur Kernphysik, Johannes Gutenberg-Universit\"at Mainz, D-55099 Mainz, Germany}

\author{P.\ Capel}
\email[Email:~]{pcapel@uni-mainz.de}
\affiliation{Institut f\"ur Kernphysik, Johannes Gutenberg-Universit\"at Mainz, D-55099 Mainz, Germany}
\affiliation{Physique Nucl\'eaire et Physique Quantique (CP 229),Universit\'e libre de Bruxelles (ULB), B-1050 Brussels, Belgium}

\begin{abstract}
We present a determination of optical potentials for $^4$He-target collisions using the double-folding method. We use chiral effective field theory nucleon-nucleon interactions at next-to-next-to-leading order combined with state-of-the-art nucleonic densities.
The imaginary part of the optical potential is obtained from the real double-folding interaction either through a proportionality constant or applying Kramers-Kronig dispersion relations.  
With these potentials, we compute the elastic scattering of $^4$He off various targets, from $^4$He to $^{120}$Sn. We study the sensitivity of our predicted cross sections to the choice of nucleon-nucleon interactions and nuclear densities. Very good agreement is obtained with existing elastic-scattering data for $^4$He energies between 100 and 400 MeV. 
\end{abstract}

\maketitle
\section{Introduction}
The interaction between colliding nuclei constitutes a significant input in the modeling of nuclear reactions~\cite{Bran97heavyion}. Given its complicated nature, the nuclear part of that interaction has typically been described by fitting parameters of phenomenological potentials, which rely on the existence of experimental data. Thus, they are accurate but lack predictive power. Recently, with the development of precise nucleon-nucleon ($NN$) interactions, efforts have been made to derive such reaction potentials from first principles. For example, in Refs.~\cite{Vora18OpPot,Idin19OpPot,Rotu20OpPot,White20OpPot}, nucleon-nucleus optical potentials have been derived using chiral effective field theory (EFT) interactions as input.

For modern nuclear forces, chiral EFT has become the standard method for developing interactions rooted in the symmetries of quantum chromodynamics (see, e.g., Refs.~\cite{Epel09RMP,Mach11PR,Hamm13RMP} for reviews). Based on a power counting scheme, $NN$ interactions can be expressed as an expansion that starts at leading order (LO), followed by contributions at next-to-leading order (NLO) and next-to-next-to leading order (N$^2$LO), etc., which provides a systematic improvement of the description of observables.

In this work, we concentrate on the derivation of nucleus-nucleus potentials through the application of the double-folding model. This technique constitutes a first-order approximation to optical potentials derived from Feshbach's reaction theory~\cite{Bran97heavyion}. In this formalism, it is possible to determine nucleus-nucleus interactions using more fundamental inputs: realistic nuclear densities and microscopic $NN$ interactions. Interesting results have been obtained in such a way to determine the real part of the optical potentials~\cite{Cham02SPdens,Pere09ImDFPot}, or the real and imaginary parts using a $G$-matrix
approach~\cite{Furu12OpPot,Mino15ChEFTCC,Khoa16DFM}.
In our previous studies~\cite{Dura17DFP,Dura20ImDisp}, we have explored the possibility to describe the interaction between two colliding nuclei through the double folding of a $NN$ interaction developed within a chiral EFT framework.

In Refs.~\cite{Dura17DFP,Dura20ImDisp}, we have studied the elastic scattering and low-energy fusion reactions of three systems: $^{12}$C-$^{12}$C, $^{16}$O-$^{16}$O, and $^{12}$C-$^{16}$O. 
To this end, we have computed optical potentials using the double-folding method with local chiral EFT potentials derived as in Refs.~\cite{Geze13QMCchi,Geze14long}. We have used these $NN$ interactions regulated in coordinate space with cutoffs $R_0=1.2$, 1.4 and 1.6~fm~\cite{Dura17DFP}. We have adopted two-parameter Fermi parametrizations~\cite{Cham02SPdens} or realistic density profiles from electron scattering~\cite{Devr87rhoel} in the folding procedure. The choice of realistic densities gives significant improvement, leading to good agreement with existing data for a variety of collision observables.

Elastic-scattering calculations are strongly sensitive to the choice of the imaginary part of the potential. To simulate this absorptive part, we have explored two possibilities: the first one is a zeroth-order approximation setting the imaginary part proportional to the real double-folding potential, as suggested in Refs.~\cite{Alv03ImPot,Pere09ImDFPot}. The second possibility is using Kramers-Kronig relations, better known in our field as \emph{dispersion relations}, which link the real and imaginary parts of the interaction~\cite{Carl89Dispersion,Gonz01DisRel}. 
Although the former way provides acceptable results, it relies on a free parameter. On the contrary, the latter approach provides an efficient constraint on the imaginary term of the nucleus-nucleus interactions, leading to much better agreement with the data without involving any free parameter.
It leads to very good agreement with elastic-scattering data at several energies for the collision of closed and non-closed shell nuclei~\cite{Dura20ImDisp}.

To further test the validity of our approach, in this work we analyze elastic scattering involving the light nucleus $^4$He, which, due to its zero spin-isospin nature and large binding energy, plays a key role in nuclear reactions as well as in nuclear astrophysics~\cite{Brow71aa,Kuku75lightN,AlGha15aN}. To this end, we start this study by analyzing the symmetric $^4$He-$^4$He collision, which is a relatively simple system from which we can draw conclusions on the interaction. We then extend our study towards much heavier targets, up to $^4$He-$^{120}$Sn. We have selected experimental data at intermediate energies to explore collisions involving nuclei for which we have reliable density profiles. 
Accordingly, we show results for the elastic-scattering of six different systems: $^4$He-$^4$He, $^4$He-$^{12}$C, $^4$He-$^{16}$O, $^4$He-$^{40}$Ca, $^4$He-$^{48}$Ca, and $^4$He-$^{120}$Sn and compare our results with experimental data~\cite{Woo854He4He,Cow944He4He,Rao004He4He,
Haus69Al104,Gils804HeCa,John034He12C,Lui014He16O,
Youn974He40Ca,Lui114He48Ca,Li104He120Sn}.
In all cases, we test the sensitivity of elastic-scattering cross sections to the $R_0$ cutoff of the $NN$ interaction, the nuclear density, as well as the impact of the description of the imaginary part. 

This paper is organized as follows: in Sec.~\ref{sec:optical} we give a brief overview of the formalism of the double-folding technique and the ways of building the imaginary part of the optical potential. In Sec.~\ref{sec:4He4He} we present results for the $^4$He-$^4$He elastic scattering. We follow in Sec.~\ref{sec:4HeX} with an analysis of results for heavier targets: $^4$He-$^{12}$C, $^4$He-$^{16}$O, $^4$He-$^{40}$Ca, $^4$He-$^{48}$Ca, and $^4$He-$^{120}$Sn. Finally, we summarize and give an outlook in Sec.~\ref{sec:sum}.

\section{Optical potentials}
\label{sec:optical}

\subsection{Real part: double-folding formalism}
The real part of the potential simulating the interaction between two nuclei can be obtained through a double-folding procedure~\cite{Satc79Folding,Bran97heavyion}.
In this formalism, the nuclear part of the potential between nucleus 1---of atomic and mass numbers $Z_1$ and $A_1$---and nucleus 2---of atomic and mass numbers $Z_2$ and $A_2$---can be constructed from a $NN$ interaction $v$ by folding it over the nucleonic densities ($\rho_1$ and $\rho_2$, respectively). For this, we follow the formalism of Refs.~\cite{Furu12OpPot,Dura17DFP}. 
The resulting antisymmetrized potential V$_\text{F}$ is the sum of a direct (D) and an exchange (Ex) contributions: $V_\text{F}=V_\text{D}+V_\text{Ex}$. The direct part is the average of the $NN$ interaction over both nucleonic densities and reads
\begin{equation}
V_\text{D}(r) = \sum_{i,j = n,p} \iint \rho^i_1({\bf r}_1) \, v^{ij}({\bf s}) \, \rho^j_2({\bf r}_2)
\, d^3{\bf r}_1 d^3{\bf r}_2 \,,
\label{eq:direct}
\end{equation}
\noindent where ${\bf r}$ is the relative coordinate between the centers of mass
of the nuclei, ${\bf r}_{1}$ and ${\bf r}_{2}$ are the inner coordinates of nucleus 1 and 2, respectively; ${\bf s}={\bf r}-{\bf r}_1+{\bf r}_2$ is the relative coordinate between any given pair of points in the projectile and target, and $\rho_{1,2}^{i}$ (with $i=n,p$) are the neutron and proton density distributions.

The exchange part of the potential accounts for the fact that, being identical, the nucleons of the projectile and the target cannot be distinguished from one another
\begin{multline}
V_\text{Ex}(r,E_\text{c.m.}) = \sum_{i,j = n,p} \iint \rho^i_1({\bf r}_1,{\bf r}_1+{\bf s})
\, v^{ij}_\text{Ex}({\bf s}) \\
\times \rho^j_2({\bf r}_2,{\bf r}_2-{\bf s}) \exp \left[\frac{i{\bf k}(r)\cdot{\bf s}}{\mu/m_N}\right] \,
d^3{\bf r}_1 d^3{\bf r}_2 \,, \label{eq:exchange}
\end{multline}
\noindent where $\mu$ is the reduced mass of the colliding
system, $v_\text{Ex}=-P_{12}v$ is the exchange contribution from the $NN$ potential, and the integral runs over the density matrices $\rho^i_{1,2}({\bf r},{\bf r} \pm {\bf s})$ of the nuclei. 

This expression leads to non-local terms in $V_\text{Ex}$. Nevertheless, the final potential can be written in local form approximating the density matrices entering in \Eq{eq:exchange} with the density matrix expansion (DME) restricted to its leading term~\cite{Nege72DME1,Bogn09DME}. As a consequence of this localization, in this channel we obtain the additional phase that renders the
double-folding potential dependent on the energy $E_\text{c.m.}$ in the center-of-mass reference frame. The momentum for the nucleus-nucleus relative motion ${\bf k}$ is related to $E_\text{c.m.}$, the nuclear part of the double-folding potential, 
and the double-folding Coulomb potential $V_\text{Coul}$ through
\begin{equation}
k^2(r)=\frac{2\mu}{\hbar^2} \, \Bigl[ E_\text{c.m.} - V_\text{F}({r},E_\text{c.m.}) - V_\text{Coul}({r}) 
\Bigr] \,. \label{eq:k}
\end{equation}
Due to the dependence of ${\bf k}$ on $V_\text{F}$, $V_\text{Ex}$ has to be determined self-consistently.

To describe the nuclear density profiles, we use densities inferred from precise electron-scattering measurements~\cite{Devr87rhoel}, and from state-of-the-art nuclear-structure calculations, such as quantum Monte Carlo (QMC) for $^4$He~\cite{Lynn17QMClight} or relativistic mean field (RMF) for heavier nuclei~\cite{Chen15NStars}.

For the $NN$ interaction $v$, we consider the potentials developed within chiral EFT in Ref.~\cite{Dura17DFP}, since they give the advantage to work in coordinate space. These potentials are regulated with cutoffs $R_0=1.2$ and $1.6$~fm, and are based on the formalism developed in Refs.~\cite{Geze13QMCchi,Geze14long}. We will show only results at N$^2$LO, which is the highest order in the chiral expansion in which these potentials are available. 
We include only two-body forces, leaving the analysis of the impact of three-body interactions for a later study.

\subsection{Imaginary part}
Between composite projectiles, a simple way to account for excitation and other inelastic processes is to consider complex interactions, known as optical potentials~\cite{Fesh58}. A general optical potential can be written as a sum of a real term independent of the energy, and a real and an imaginary terms dependent on the energy:
\begin{equation}
U_F(r,E_\text{c.m.})=V_\text{D}(r)+V_\text{Ex}(r,E_\text{c.m.})+iW(r,E_\text{c.m.})\,,
\label{eq:U_F}
\end{equation}
The resulting double-folding potential sum of Eqs.~(\ref{eq:direct}) and~(\ref{eq:exchange}) is purely real. A simple way to determine the imaginary part $W$ is to assume it proportional to the real double-folding potential ~\cite{Alv03ImPot,Pere09ImDFPot},
\begin{equation}
 W=N_W V_F  \,,
\label{eq:NW}
\end{equation} 
where $N_W$ is a constant, which we take in the range 0.6--0.8~\cite{Pere09ImDFPot,Dura17DFP}.
Alternatively, we find a much better agreement with data when the Kramers-Kronig relations~\cite{Kron26Disp,Kram27Disp,Carl89Dispersion} are used to construct $W$ from $V_\text{F}$~\cite{Dura20ImDisp}.
Writing the local complex optical potential $U$ between two nuclei as Eq.~(\ref{eq:U_F}), its imaginary part can be calculated through:
\begin{equation}
W(r,E_\text{c.m.})=-\frac{1}{\pi}\mathcal{P}\int_{-\infty}^{+\infty} \frac{V_\text{Ex}(r,E)}{E-E_\text{c.m.}}dE\,,
\label{eq:W_Disp}
\end{equation}
\noindent where $\mathcal{P}$ represents the principal value integral.

These relations are the application of the Sokhotski-Plemelj theorem assuming that the function $U$ is analytical, holomorphic, and square integrable in the upper half of the complex plane.  
The application of the link between real and imaginary parts of a function was born in the field of Optics~\cite{Kron26Disp,Kram27Disp}, and the idea was adapted to several physical problems, including nuclear reactions~\cite{Fesh62,Pass67DispRel,Maha86DispRel}. The initial derivation of the Kramers-Kronig relations was obtained for time-dependent functions. Interactions that include levels beyond Hartree-Fock give energy contributions that correspond to a time-dependence of the wave function. However, it has been proven to hold also for spatial variables~\cite{Hors15SpatKramKro}. In our case, the energy-dependence of our optical potential arises from spatial nonlocality, since it is a consequence of the antisymmetrization of particles. In this sense, our application of the Kramers-Kronig relations is not fully equivalent to the dispersion relations derived by Feshbach~\cite{Fesh62}, but it is mathematically valid and presents a first parameter-free derivation of the optical potential.
Furthermore, at energies above the Coulomb barrier, the imaginary part has important contributions from the energy dependence that arises from the exchange term~\cite{Maha86DispRel}, and we expect this approach to give a good first-order contribution to the imaginary part. 

\section{$^4\text{He}$-$^4\text{He}$ elastic scattering}
\label{sec:4He4He}

\subsection{$^4\text{He}$-$^4\text{He}$ potential}
To start this study, we analyze the elastic scattering of the symmetric system $^4\text{He}$-$^4\text{He}$ at two laboratory energies: 198.8 and 280~MeV, which correspond to the experimental conditions of Refs.~\cite{Woo854He4He,Cow944He4He,Rao004He4He}. 

\begin{figure}[tb]
\begin{center}
\includegraphics[width=0.99\columnwidth]{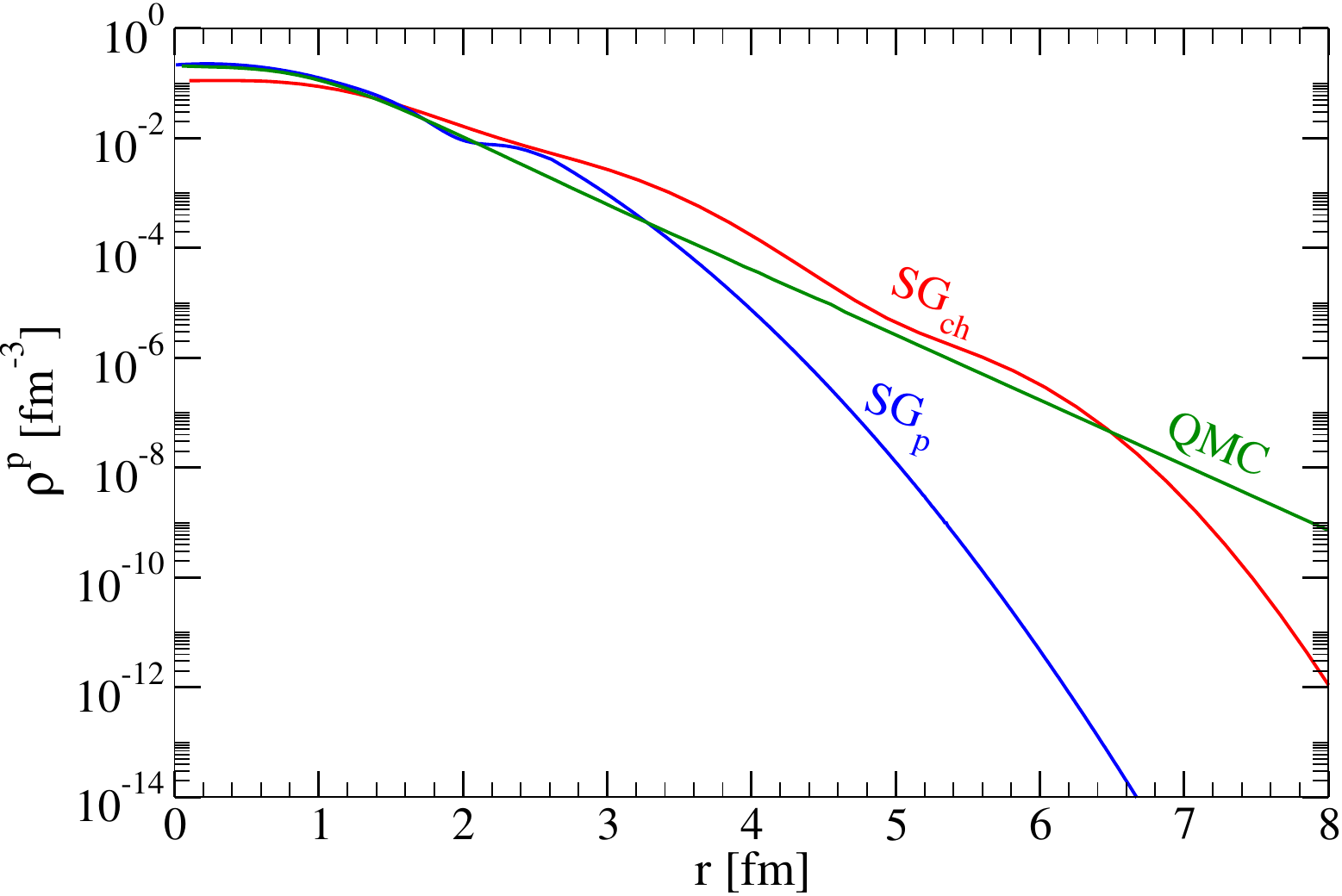}
\caption{Proton density profiles for $^{4}$He: the Sum-of-Gaussians charge density derived from electron scattering~\cite{Devr87rhoel} (SG$_\text{ch}$), the corresponding proton particle density (SG$_\text{p}$), and the density calculated using QMC with local N$^2$LO potentials~\cite{Lynn17QMClight}.} 
\label{fig:4He_rho}
\end{center}
\end{figure}

Since $^4$He is a light and stable nucleus with an equal number of protons and neutrons, we make the approximation $\rho^{p}=\rho^{n}$. To describe the proton density, we consider three possibilities: a charge density obtained through electron scattering and parametrised as a sum of Gaussians in Ref.~\cite{Devr87rhoel} (denoted as SG$_\text{ch}$); the corresponding nucleonic density, obtained through the numerical inverse Fourier transform of the Fourier transformed SG$_\text{ch}$ divided by the nucleonic form factor~\cite{Mohr15CODATA,Papo15neutr} (named SG$_\text{p}$); and finally, we also use a proton density profile computed within the Quantum Monte Carlo framework (QMC) with two- and three-body local chiral interactions at N$^2$LO with cutoff $R_0=1.2$ fm~\cite{Geze14long,Lynn17QMClight}. The two-body part of this interaction is the same as the one we use for the calculation of the nucleus-nucleus potentials.
These density profiles can be seen in Fig.~\ref{fig:4He_rho}: SG$_\text{ch}$ in red, SG$_\text{p}$ in blue, and QMC in green. 
The SG$_\text{p}$ and QMC density profiles have a similar shape up to $r\approx 3$ fm, while SG$_\text{ch}$ has a more diffuse behavior: a density lower inside the nucleus, and higher at its surface (in the region 2~fm~$\lesssim r\lesssim5$~fm). The unrealistic behavior of SG$_\text{ch}$ at large distance is due to the nature of its parametrisation with Gaussian functions.

Using the Kramers-Kronig relations to obtain the imaginary part of the $^4\text{He}$-$^4\text{He}$ potential, we have observed, as in Ref.~\cite{Dura20ImDisp}, that the exchange term (\ref{eq:exchange}) shows the same $r$-dependence at all energies for all the different $NN$ interactions and densities considered. Accordingly, we can write in this case:
\begin{equation}
V_\text{Ex}(r,E)=f_\text{Ex}(r)V^0_\text{Ex}(E)\,,
\end{equation}
\noindent and the imaginary part can be calculated as, 
\begin{equation}
W(r,E_\text{c.m.})=-\frac{ f_\text{Ex}(r)}{\pi}\mathcal{P}\int dE\frac{V^0_\text{Ex}(E)}{E-E_\text{c.m.}}\,.
\end{equation}
The advantage of this expression compared to Eq.~(\ref{eq:W_Disp}) is that it suffices to compute the value of the potentials at different energies at one given $r$, that we can choose arbitrarily, to obtain the energy dependence and perform the integration (\ref{eq:W_Disp}). This integral requires the depth of $V_\text{Ex}$ at negative energies, that we set as $V^0_\text{Ex}(E<0)=0$. Note that we have tested that setting it to the value of $V^0_\text{Ex}(E=0)$ instead does not change the results at the energy range of interest in this study.

\subsection{Elastic-scattering cross sections}

\begin{figure}[tb]
\begin{center}
\includegraphics[width=0.99\columnwidth]{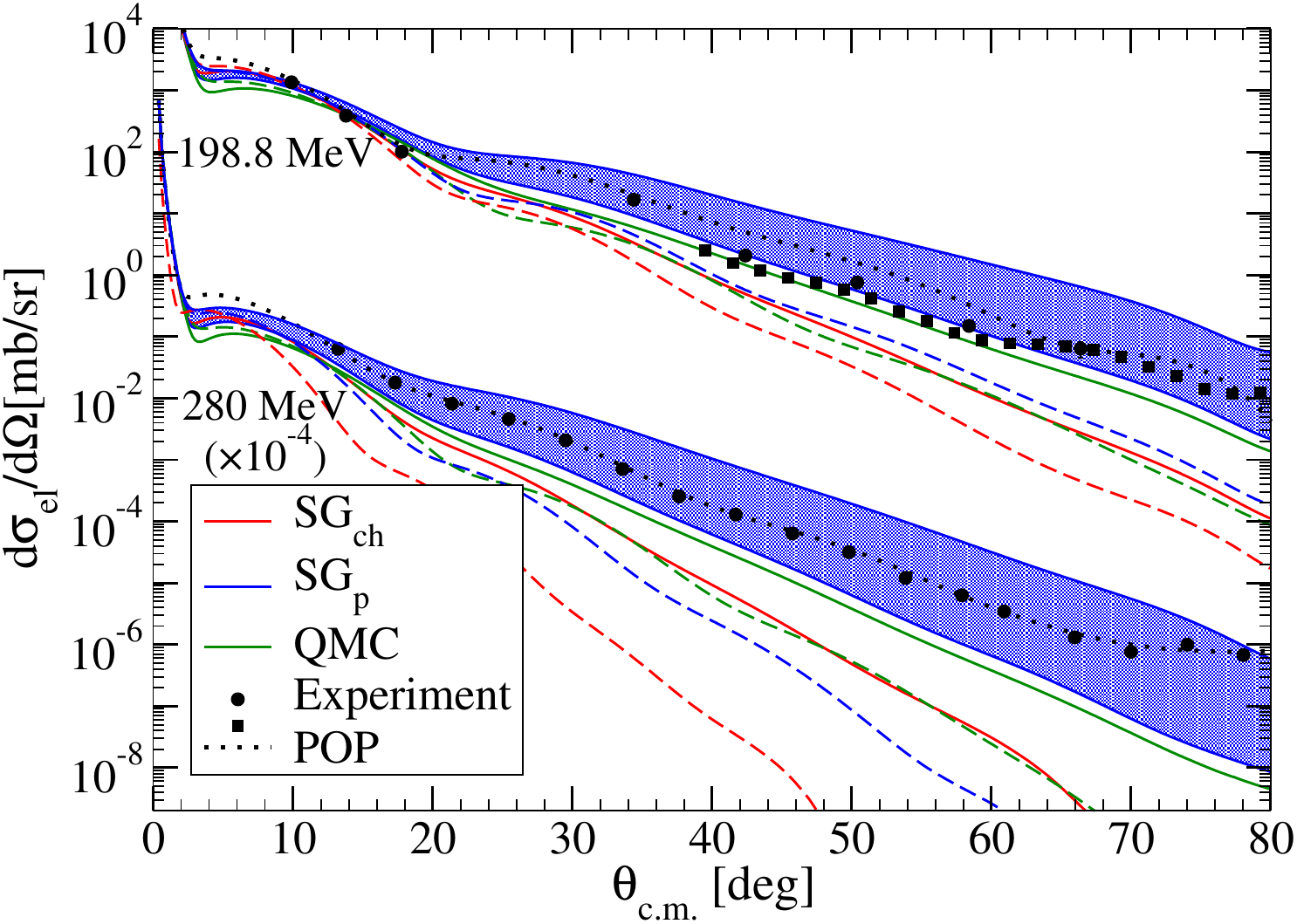}
\caption{Elastic scattering cross sections for $^4$He-$^{4}$He at $E_\alpha=198.8$, and 280~MeV. The imaginary part is obtained through Kramers-Kronig relations, Eq.~(\ref{eq:W_Disp}), (solid lines) or considered to be $W=0.6\,V_F$, Eq.~(\ref{eq:NW}), (dashed lines). For $R_0=1.2$~fm, results with SG$_\text{ch}$, SG$_\text{p}$ and QMC are shown in red (lower lines), blue (upper lines/bands), and green (middle lines), respectively. For the SG$_\text{p}$, the blue band corresponds to a change in $R_0$ from 1.2~fm to 1.6~fm. POP from Refs.~\cite{Woo854He4He,Rao004He4He} are shown for comparison as black dotted lines. Data from Refs.~\cite{Woo854He4He,Cow944He4He,Rao004He4He}.} 
\label{fig:4He4He_el}
\end{center}
\end{figure}

Figure~\ref{fig:4He4He_el} shows the results for elastic-scattering cross sections at $E_\alpha=198.8$, and 280 MeV~\cite{Woo854He4He,Cow944He4He,Rao004He4He}. 
The color code matches that of Fig.~\ref{fig:4He_rho} for the different density profiles: SG$_\text{ch}$ in red, SG$_\text{p}$ in blue, and QMC in green. To illustrate the impact of the $NN$ cutoff, we show the blue bands, which reflect the $R_0$ variation between 1.2 and 1.6~fm for the SG$_\text{p}$ density. For all densities, the solid lines depict the results obtained with Kramers-Kronig relations. The blue and green curves are close to each other, indicating that the double-folding process probes the densities up to $r\simeq3$~fm, since the SG$_\text{p}$ and QMC profiles in Fig.~\ref{fig:4He4He_el} have different tails but lead to similar cross sections. The cross sections obtained with SG$_\text{ch}$, which is the most diffuse density, drop faster with the angle and do not reproduce data as well as the other two, even though this is also a sensible density choice. 

To compare these results with the cruder approximation [Eq.~(\ref{eq:NW})], we show with  dashed lines the results obtained with $N_W=0.6$, which is the value recommended in Ref.~\cite{Pere09ImDFPot}. These results confirm our previous observations that Kramers-Kronig relations give better reproduction of the data, especially at larger angles~\cite{Dura20ImDisp}.

For both energies, the blue band gives very good agreement with experiment and also with the phenomenological optical potentials (POP) from Ref.~\cite{Woo854He4He,Rao004He4He} (black dotted lines). These potentials consist of two Woods-Saxons parametrizations with a total of 6 fitted parameters. The increase in the width of the bands reflects the fact that at large angles the short-range $NN$ physics becomes more relevant.
Let us stress that our results using Kramers-Kronig relations to determine the imaginary potentials are obtained without any fitting parameter.

\section{$^4\text{He}$ elastic scattering off heavier targets}
\label{sec:4HeX}

\subsection{Kramers-Kronig relations in asymmetric cases}

We consider now the elastic scattering of $^4$He with five different targets: $^{12}$C, $^{16}$O, $^{40}$Ca, $^{48}$Ca, and $^{120}$Sn for which we have access to precise nucleonic densities~\cite{Devr87rhoel,Chen15NStars} and experimental data sets to which to compare~\cite{Haus69Al104,Gils804HeCa,John034He12C,Lui014He16O,Youn974He40Ca,Lui114He48Ca,Li104He120Sn}. Interestingly, this time the $r$-dependence of $V_\text{Ex}$ varies with the collision energy. This is different from what has been seen in Sec.~\ref{sec:4He4He} and in our previous study~\cite{Dura20ImDisp}, probably due to the significant asymmetry between the projectile and the target. Therefore, the energy dependence affects both the depth of the potential and its radial shape. To apply the Kramers-Kronig relations, we need to use Eq.~(\ref{eq:W_Disp}) and perform the energy integration of the exchange potential at each radial point for all energies. The resulting imaginary part thus exhibits a radial dependence different from that of $V_\text{Ex}$.

\subsection{Scattering on $^{40}$Ca}

Let us first present and detail the results for $^4$He-$^{40}$Ca because it best illustrates the issues at hand in these calculations. In this section, we extract conclusions that are valid also for the other systems we will discuss in Sec.~\ref{sec:heavier}. As in Sec.~\ref{sec:4He4He}, we consider SG$_\text{ch}$, SG$_\text{p}$, and QMC as density profiles for $^4$He. For $^{40}$Ca, we take the sum-of-Gaussians parametrization of the charge density inferred from the elastic scattering of electrons from Ref~\cite{Devr87rhoel} (SG$_\text{ch}$), and the corresponding nucleonic density obtained through the Fourier transform of SG$_\text{ch}$ divided by the nucleonic form factor (SG$_\text{p}$). In these two cases, we assume the approximation $\rho^{p}=\rho^{n}$. We consider also a density profile obtained with RMF calculations~\cite{Chen15NStars}, which provides estimates for $\rho^{p}$ and $\rho^{n}$. 

\begin{figure}[tb]
\begin{center}
\includegraphics[width=0.99\columnwidth]{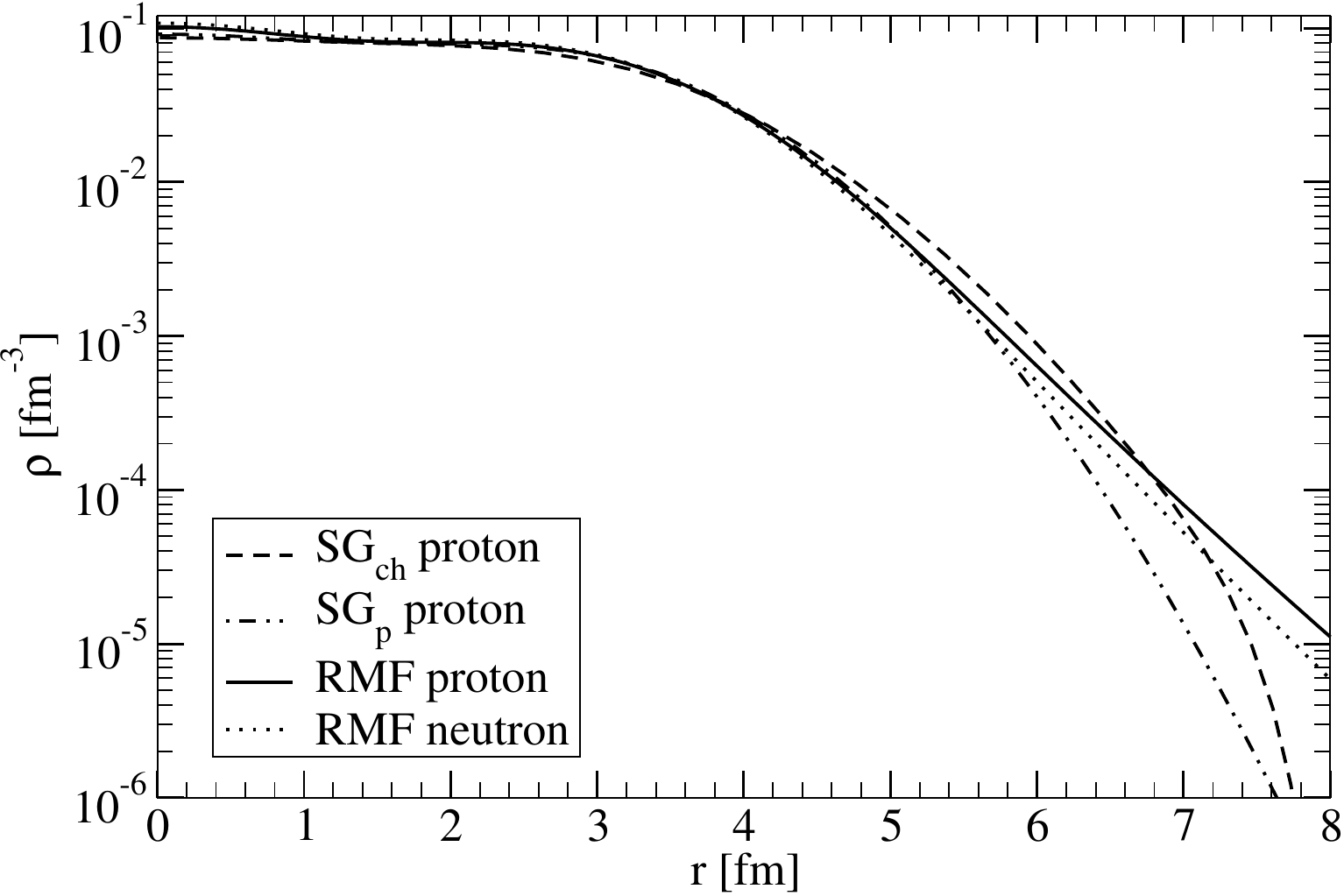}
\caption{$^{40}$Ca density profiles:
charge density obtained from electron scattering~\cite{Devr87rhoel} (SG$_\text{ch}$, dashed line); the corresponding proton density (SG$_\text{p}$, dashed-dotted line), and RMF calculation for protons (solid line) and neutrons (dotted line)~\cite{Chen15NStars}.} 
\label{fig:40Ca_dens}
\end{center}
\end{figure}

Figure~\ref{fig:40Ca_dens} shows the density profiles for $^{40}$Ca. As it was the case for $^4$He, SG$_\text{ch}$ (dashed line) gives the most diffuse density profile up to $r\simeq7$~fm where the Gaussian parametrization leads to an unrealistic drop. Once again, the SG$_\text{p}$ and RMF profiles (dashed-dotted and solid lines, respectively) show similar behavior between $r\approx 1$ fm and $r\approx 6$~fm for the proton distribution. We show the RMF neutron density as the dotted line. It can be seen that these proton and neutron densities (solid and dotted lines) are close to each other, justifying the approximation assumed for SG$_\text{ch}$ and SG$_\text{p}$. 

\begin{figure}[tb]
\begin{center}
\includegraphics[width=0.99\columnwidth]{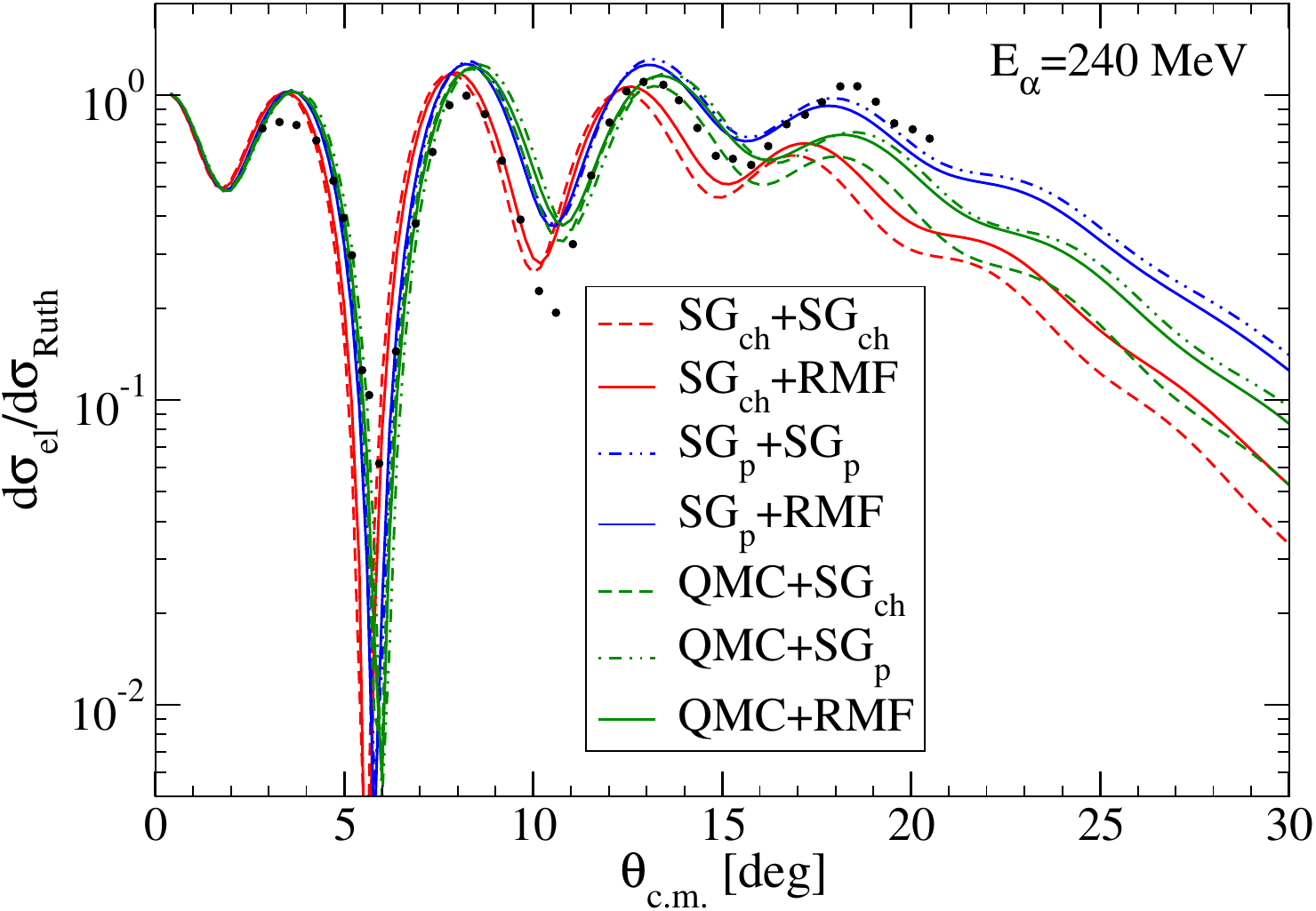}
\caption{Elastic-scattering cross sections (normalized to Rutherford) for $^4$He-$^{40}$Ca at $E_\alpha=240$ MeV. The potentials are calculated with $R_0=1.2$ fm, and the imaginary part is obtained through Kramers-Kronig relations. Results obtained with $\rho_{^{40}\text{Ca}}$ SG$_\text{ch}$, SG$_\text{p}$ or RMF are shown respectively with dashed, dashed-dotted, and solid lines. For each case, results obtained with $\rho_{^4\text{He}}$ described as SGch, SGp or QMC are shown as red (lower lines), blue (upper lines), and green (middle lines), respectively. Experimental data taken from Ref.~\cite{Youn974He40Ca}.} 
\label{fig:4He40Ca_el_disp}
\end{center}
\end{figure}

Figure~\ref{fig:4He40Ca_el_disp} shows elastic-scattering cross sections normalized to Rutherford for $^4$He-$^{40}$Ca at $E_\alpha=240$ MeV. These results illustrate the sensitivity of our calculations to the choice of the density of both the projectile and the target. In this figure and in the following ones, the labels design the density profiles chosen for $^4$He and the target, in that order. For example, ``SG$_\text{ch}$+RMF'' means $\rho_{^4\text{He}}=\rho_\text{SGch}$ and $\rho_{^{40}\text{Ca}}=\rho_\text{RMF}$.
All these results are calculated with the $NN$ cutoff $R_0=1.2$ fm, and the imaginary part is obtained through Kramers-Kronig relations. Cross sections obtained with $\rho_{^4\text{He}}$ described as SG$_\text{ch}$, SG$_\text{p}$ or QMC are shown as red, blue, and green lines, respectively (following the color code of Fig.~\ref{fig:4He_rho}). Results obtained with $\rho_{^{40}\text{Ca}}$ SG$_\text{ch}$, SG$_\text{p}$ or RMF are shown, respectively, with dashed, dashed-dotted, and solid lines (following the line types in Fig.~\ref{fig:40Ca_dens}). From this figure, we can conclude that the density of $^4\text{He}$ has the most significant impact on the results, since the curves are grouped by colors. As it was seen in the case of $^4$He-$^{4}$He, the results obtained with SG$_\text{ch}$ (most diffuse density) are the most forward focussed, even in the first minimum, while results with QMC are shifted towards larger angles, starting from the second minimum. SG$_\text{p}$ gives cross sections that are in phase and show the best agreement with experimental data~\cite{Youn974He40Ca}. Nevertheless, it is important to note that using this $NN$ interaction no density choice enables us to correctly reproduce the second minimum in the data. In our calculations, we found that using $N_W$ to describe the imaginary part also leads to results that are dominated by $\rho_{^4\text{He}}$ and show the same kind of behavior seen in Fig.~\ref{fig:4He40Ca_el_disp}.

As it was the case in $^4$He-$^4$He scattering, we find that Kramers-Kronig relations reproduce data at large angles. We want to remind the reader that in the calculation of these cross sections there is no parameter fitting. Using Kramers-Kronig relations with SG$_\text{p}$ for $^4$He overestimates the magnitude of the data between 8$^{\circ}$ and 16$^{\circ}$, but gives the right magnitude at large angles and leads to the right oscillatory pattern when compared to data. 

Note that we have explored a fourth density profile for $^{40}$Ca obtained through Coupled Cluster calculations using N$^2$LO$_\text{sat}$ potentials~\cite{Hage16NatPhys}. These calculations give a similar density profile to the RMF density around the surface area, and lead to practically indistinguishable elastic-scattering cross sections. This shows that such an observable is not sensitive enough to distinguish the differences between precise nuclear-structure calculations of the density. 

\subsection{Results for heavier targets}
\label{sec:heavier}

\subsubsection{Medium-mass nuclei}

\begin{figure*}[]
\begin{center}
\includegraphics[width=0.97\columnwidth]{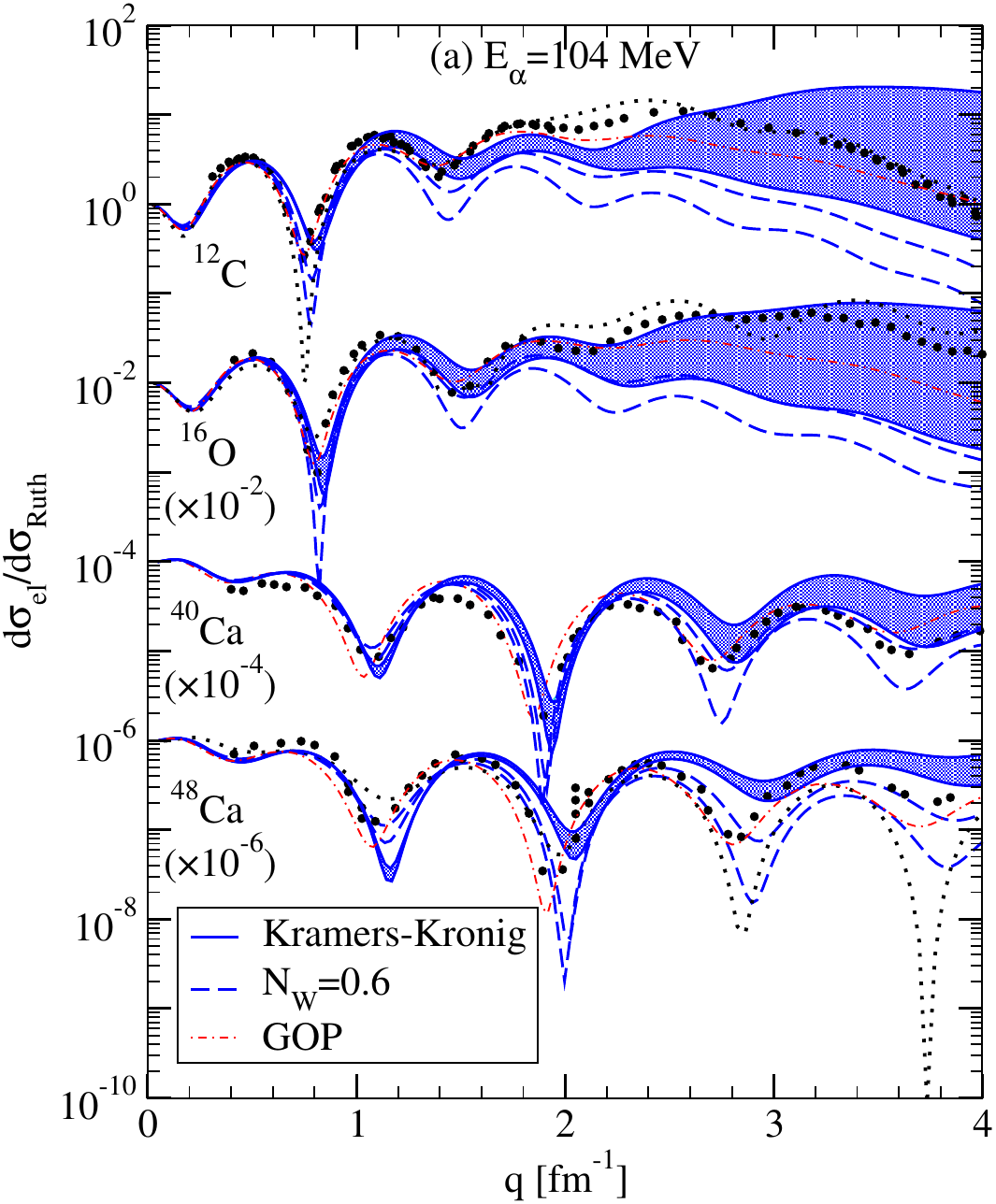}
\includegraphics[width=0.97\columnwidth]{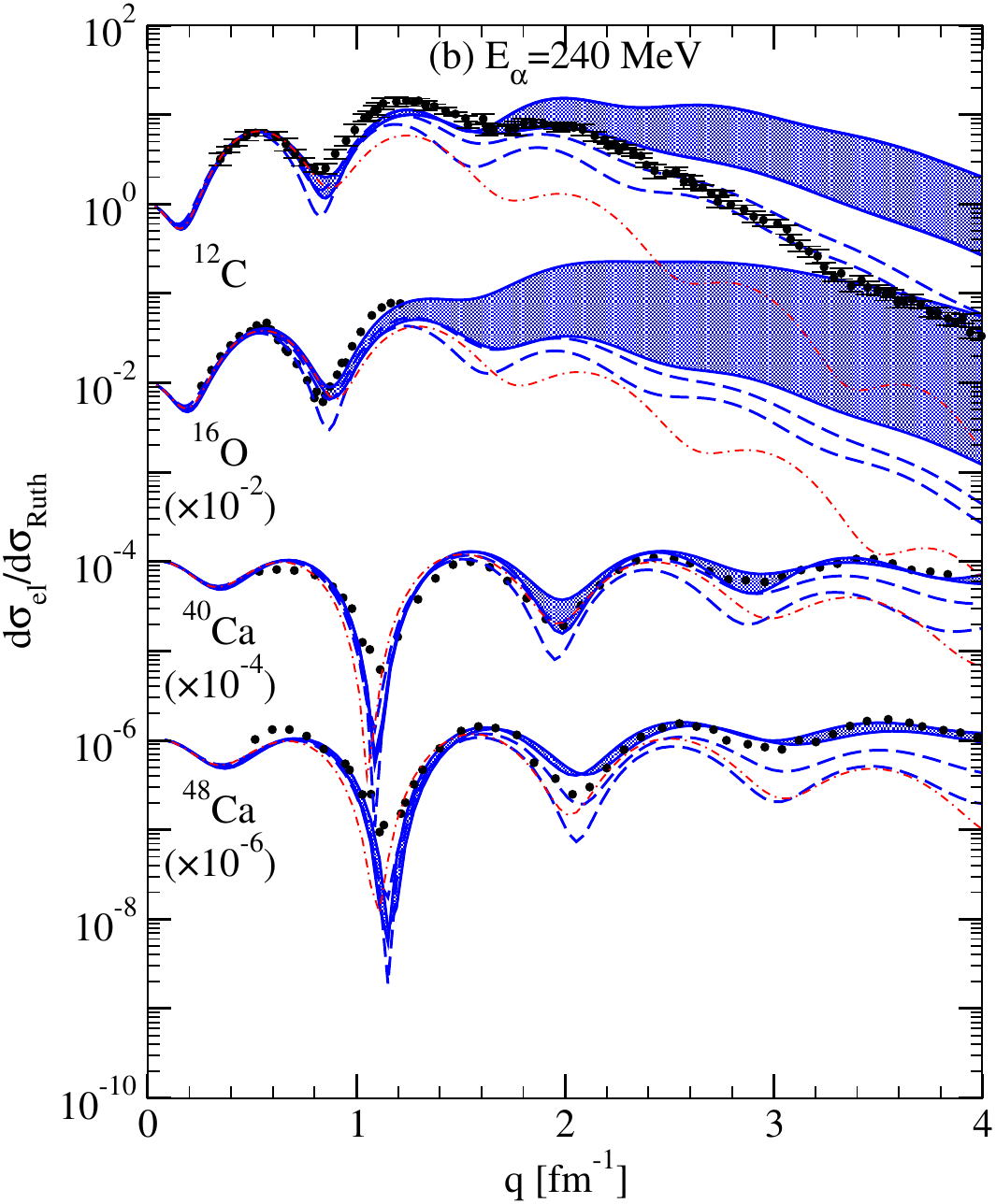}
\caption{Elastic scattering cross sections (normalized to Rutherford) as a function of the momentum transfer $q$ for $^4$He-$^{12}$C, $^4$He-$^{16}$O, $^4$He-$^{40}$Ca and$^4$He-$^{48}$Ca at (a) $E_\alpha=104$ and (b) 240 MeV. The bands show the $R_0$ dependence. The imaginary part was obtained through Kramers-Kronig relations (solid lines) or considered to be $W=0.6\,V_F$ (dashed lines). The chosen densities correspond to the combination that best reproduces experimental data taken from~\cite{Haus69Al104,Gils804HeCa}, and~\cite{John034He12C,Lui014He16O,Youn974He40Ca,Lui114He48Ca}. For comparison, results with POP~\cite{Haus69Al104,Youn974He40Ca} are shown as black dotted lines. Cross sections obtained with the GOP of Ref.~\cite{Nolt87GOP} are shown as red dash-dotted lines.} 
\label{fig:4HeX_el_q}
\end{center}
\end{figure*}

We have observed that the behavior seen in Fig.~\ref{fig:4He40Ca_el_disp} is general for elastic scattering of the form $^4$He(X,X)$^4$He, where X denotes a target heavier than $^4$He. Figure~\ref{fig:4HeX_el_q} shows results for the cross section as a function of the momentum transfer $q$, for $^4$He impinging on $^{12}$C, $^{16}$O, $^{40}$Ca, and $^{48}$Ca at (a) $E_\alpha=104$ and (b) 240 MeV. All the shown cross sections are calculated choosing SG$_\text{p}$ to describe the $^4$He density, because, as illustrated in Figs.~\ref{fig:4He4He_el} and~\ref{fig:4He40Ca_el_disp}, it gives better results in general. In the case of the $Z=N$ targets, we also chose SG$_\text{p}$~\cite{Devr87rhoel}, while for $^{48}$Ca the density is taken to be that from RMF calculations~\cite{Chen15NStars}. 
The shaded bands show the $R_0$ 1.2--1.6 fm dependence of the cross sections obtained using Kramers-Kronig relations to constrain the imaginary part of the potentials. 
To further illustrate the validity of our approach, the dashed lines show the cross sections calculated with $N_W=0.6$. 
We also show results obtained with $\alpha$-nucleus global optical potentials (GOP) from Ref.~\cite{Nolt87GOP} (red dash-dotted lines), which have not been fitted to the data sets that are analyzed in this work. Finally, for $^{12}$C and $^{16}$O at 104 MeV, the black dotted lines depict the cross sections calculated with POPs from Ref.~\cite{Haus69Al104}, which are modeled as ``wine-bottle" potentials. These phenomenological potentials use 6 parameters and are fitted to the corresponding data sets.

The agreement with the data is generally good for all targets. At small angles, Kramers-Kronig relations and the choice $N_W=0.6$ give comparably good results. However, Kramers-Kronig relations are necessary to reproduce the large-angle behavior of the data, since the results obtained with $W=0.6\, V_F$ deviate from experiment for large $q$. 
The sensitivity to $R_0$ is large for light targets and decreases with the target mass, and, in general, experimental data lies within that uncertainty band. At 104 MeV, our results for $^{12}$C and $^{16}$O suffer in comparison to those obtained with POP, which were fitted for each system and energy. Compared to the GOP of Ref.~\cite{Nolt87GOP}, our agreement with data is as good or even better in some cases at both energies and for all targets. 
We want to point out that for $^4$He-$^{40}$Ca at 240 MeV, the cross section corresponding to $R_0=1.6$ fm (lower line) reproduces the second minimum without increasing the uncertainty of our results in the first minimum, which is an indication of the sensitivity of our problem to short-range physics.
It is important to note that we show the results for SG$_\text{p}$+RMF for $^{48}$Ca at both energies for consistency, however the best results at $E_\alpha=104$ MeV are obtained with SG$_\text{ch}$+RMF. Since this is an exception to what we have observed in the other cases, we take this to be an accidental result.

\subsubsection{$^{120}$Sn}
\begin{figure}[tb]
\begin{center}
\includegraphics[width=0.99\columnwidth]{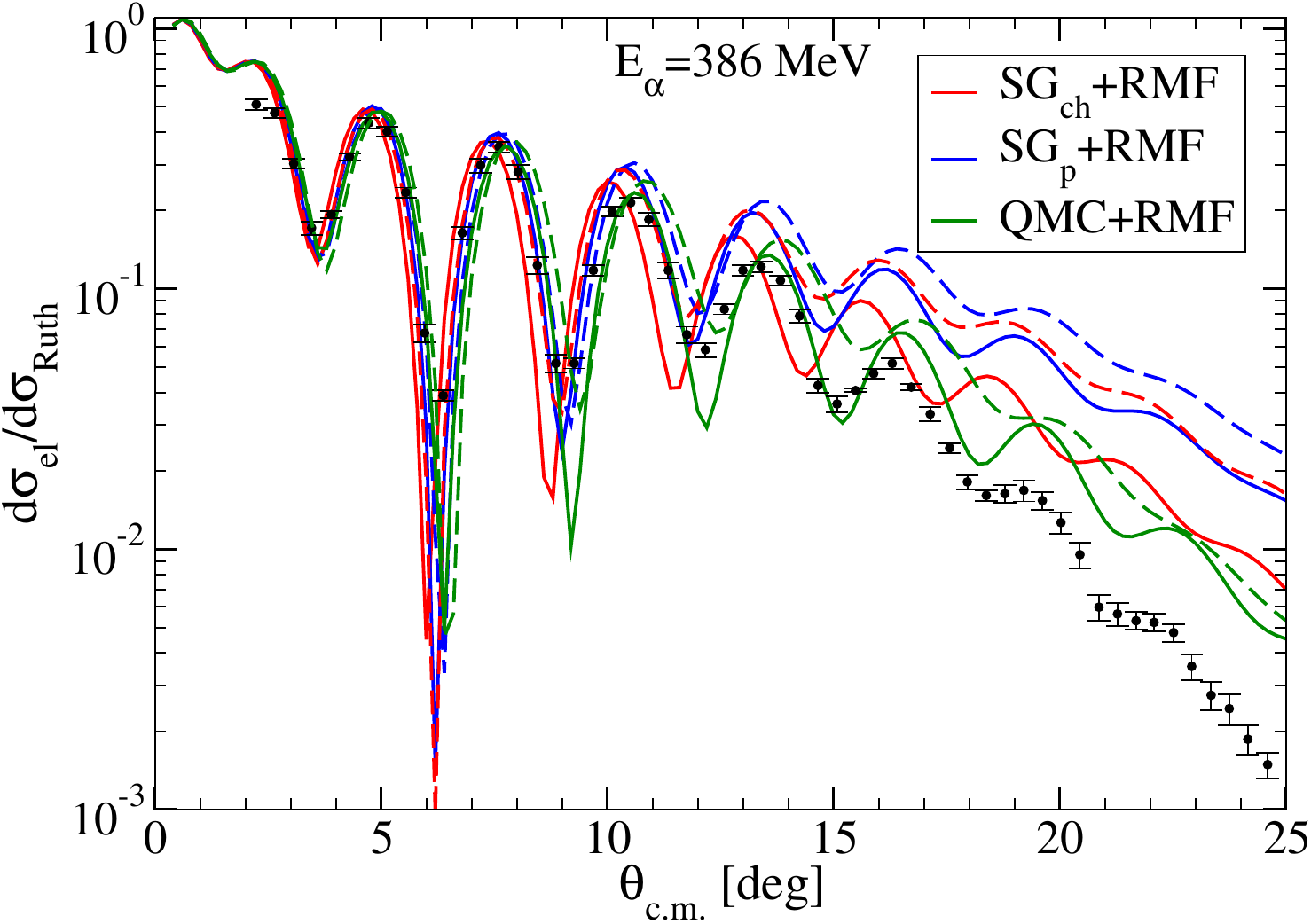}
\caption{Elastic-scattering cross sections (normalized to Rutherford) as a function of the center-of-mass angle for $^4$He-$^{120}$Sn at $E_\alpha=386$ MeV using different $^4$He densities for $R_0=1.2$ fm. The imaginary part is obtained through Kramers-Kronig relations (solid lines) or considered to be $W=0.6 \,V_F$ (dashed lines). For each case, results obtained with $\rho_{^4\text{He}}$ described as SGch, SGp or QMC are shown as red (middle lines), blue (upper lines), and green (lower lines), respectively. Experimental data from Ref.~\cite{Li104He120Sn}.} 
\label{fig:4He120Sn_el}
\end{center}
\end{figure}

Another application of optical potentials generated by double folding can be found in Fig.~\ref{fig:4He120Sn_el}, that displays cross sections for $^4$He-$^{120}$Sn elastic scattering at $E_\alpha=386\,\text{MeV}$~\cite{Li104He120Sn}. The imaginary part is obtained through Kramers-Kronig relations (solid lines) with $N_W=0.6$ (dashed lines). As for the other targets, we observe a significant sensitivity of our calculations to the choice of the $^4$He density, mostly at large angles. Once again, SG$_\text{ch}$ (red lines) leads to cross sections that are shifted towards forward angle, SG$_\text{p}$ (blue lines) produces a cross section mostly in phase with the experimental data~\cite{Li104He120Sn}, and QMC (green lines) is slightly shifted towards larger angles starting at around 10$^{\circ} $. However, the magnitude of the cross section is closer to experimental data using the QMC density, and this choice provides the best overall description of the experimental cross section. For this system, we have also tested two more $^{120}$Sn densities: a two-parameter Fermi expression~\cite{Cham02SPdens}, and a three-parameter Gaussian~\cite{Devr87rhoel} profile obtained from electron scattering. We only depict results with RMF densities~\cite{Chen15NStars} for $^{120}$Sn, since our observations using the other profiles have shown that the choice of this density has little influence on the cross sections.

\begin{figure}[tb]
\begin{center}
\includegraphics[width=0.99\columnwidth]{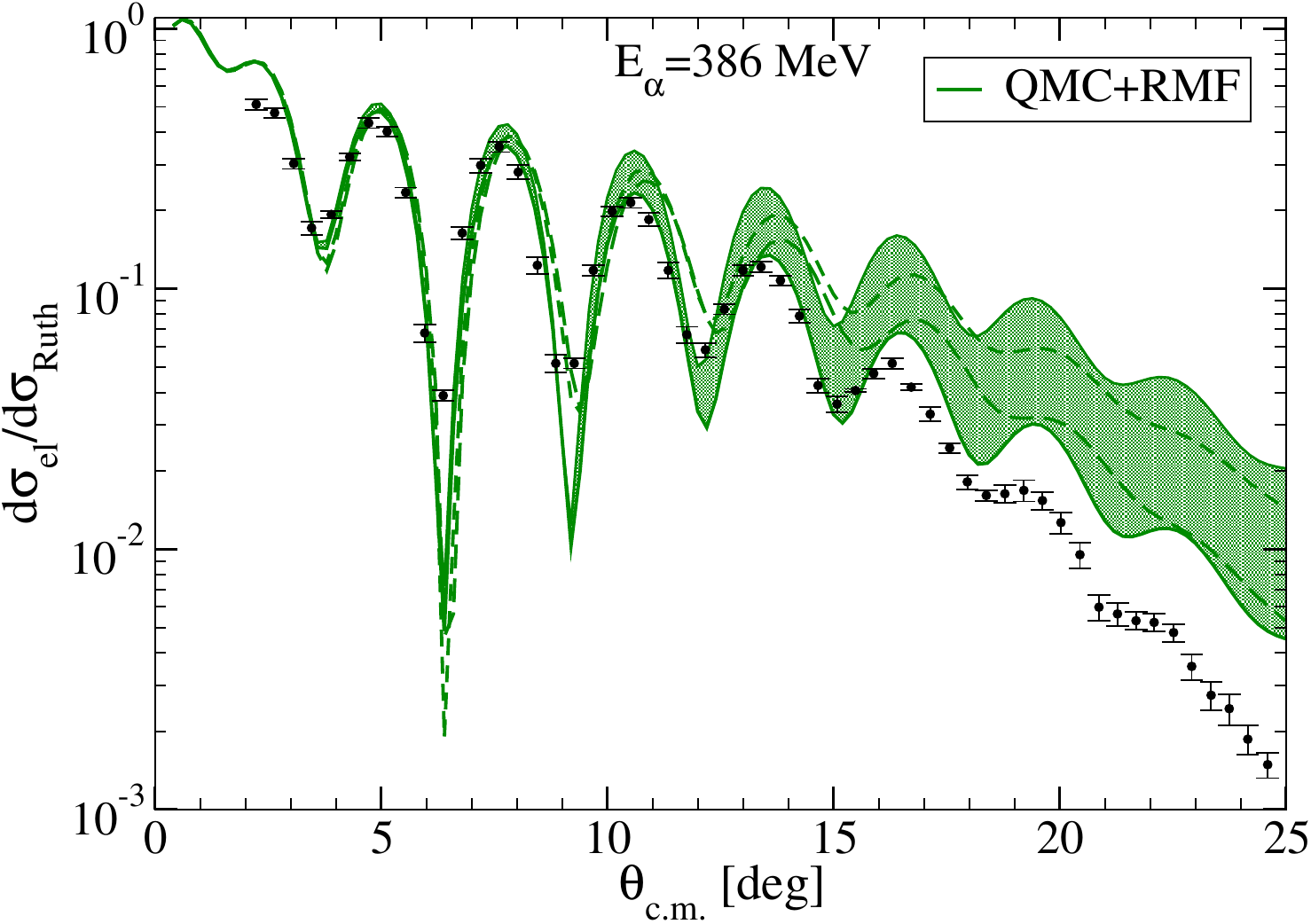}
\caption{Same as Fig.~\ref{fig:4He120Sn_el} exploring the $R_0$ dependence for QMC+RMF.} 
\label{fig:4He120Sn_el_RMF}
\end{center}
\end{figure}

Figure~\ref{fig:4He120Sn_el_RMF} explores the $R_0$ dependence of the cross sections using the density combination QMC+RMF. The green band gives the $R_0=1.2$--1.6 fm sensitivity when using Kramers-Kronig relations to calculate the imaginary part (dashed lines show this sensitivity using $N_W=0.6$). At large angles, the oscillatory pattern is better reproduced by the Kramers-Kronig relations, even though the results are similar in magnitude. Note that, in this case, setting $N_W=0.8$ (as explored in our previous works~\cite{Dura17DFP,Dura20ImDisp}) would give a better description of the data. This need for more absorption is probably due to the lower excitation energy of $^{120}$Sn, as well as its higher density of possible excited states compared to the other targets considered here. In order to better describe this reaction, one should account for the spectrum of $^{120}$Sn, that would lead to a new source of energy dependence which describes channels that our model does not contemplate. 

\section{Conclusions and Outlook}
\label{sec:sum}
We have presented the derivation of $\alpha$-target optical potentials through the double folding of local chiral EFT $NN$ potentials~\cite{Geze13QMCchi,Geze14long} over realistic nucleonic densities. To calculate their imaginary part, a zeroth-order solution is to assume it to be proportional to the real double-folding potential~\cite{Alv03ImPot}, which gives an interaction that depends on the proportionality parameter $N_W$. The approach we have mostly explored in this work is to use the Kramers-Kronig relations, which gives us a first derivation of potentials generated from first principles without any fitting parameter. Even though we do not include energy contributions from coupling to excited states, at sufficiently high energy this approximation is justified, and leads to cross sections that are in fair agreement with experimental data.
Within this framework, we were able to reproduce elastic-scattering data for the collision on $^4$He, $^{12}$C, $^{16}$O, $^{40}$Ca, $^{48}$Ca and $^{120}$Sn between 100 and 400 MeV.

Kramers-Kronig relations lead to a good prediction of data at large angles. This is true for all targets, but especially clear for $^{12}$C, and less good for $^{120}$Sn, target for which our simplified approach might not be sufficient.
Contrary to what we have observed in Ref.~\cite{Dura20ImDisp}, for asymmetric scattering, the radial shape of the imaginary part is no longer identical to that of the exchange part of the real potential. In these cases, to apply the Kramers-Kronig relations, the integration over the energy must be performed for each radial point.

We have seen that there are two major inputs for these calculations: the $\alpha$ density and the $NN$ interaction. Both of them affect significantly the angular dependence of the cross section.
We have also observed that the impact of the target density is less significant for heavier nuclei.

As in our previous work, we find that the use of realistic density profiles combined with Kramers-Kronig relations is a necessary first step towards a better description of the imaginary part of nucleus-nucleus potentials, this stays valid for non-symmetric systems and heavy nuclei. There remain several paths for improvement, at the level of both the many-body folding method and the input interactions. 
First, it would be interesting to study the impact of going beyond leading order in the DME used in Eq.~(\ref{eq:exchange}). It is also necessary to determine the impact of a calculation beyond Hartree-Fock and the nonlocal contributions that
would arise (see, e.g., Refs.~\cite{Fesh58,Fesh62}). Accounting for the excited spectrum of the colliding nuclei would also refine the description of the imaginary part of the optical potential through the application of dispersion relations to these energy-dependent terms. This would further improve our potentials and their description of the scattering processes, especially at low collision energies~\cite{Maha86DispRel}. Finally, the role of three-nucleon interactions needs to be investigated in this approach, as they also enter at N$^2$LO. We have observed in preliminary calculations that the contribution to the nucleus-nucleus potential $^{16}$O-$^{16}$O arising from three-nucleon interaction is very small compared to the two-body contributions discussed here~\cite{Webe17TFP}, but the role of these interactions should be investigated in different systems.

Knowing that there is no fitting or scaling parameter in our framework, we can conclude that there is excellent agreement between our calculations of elastic-scattering cross sections and experimental data. 
These results hint strongly towards the interest of studying the impact of using density profiles based also on chiral EFT interactions to analyse the results within a fully consistent model that would bridge reactions and structure.

\section*{Acknowledgments}
We thank J.\ Lynn for providing the QMC densities, J.\ Piekariewicz for the RMF density profiles, and S.\ Bacca for the Coupled-cluster densities. We also thank the International Atomic Energy Agency that provided the experimental data through their web page \href{www-nds.iaea.org}{www-nds.iaea.org}. 
This work was supported by the PRISMA+ (Precision Physics, Fundamental Interactions and Structure of Matter) Cluster of Excellence, the European Union's Horizon 2020 research and innovation programme under Grant Agreement No. 654002, and 
Deutsche Forschungsgemeinschaft (DFG, German Research Foundation) -- Projekt-ID 204404729 -- SFB 1044 and Projekt-ID 279384907 -- SFB 1245.

\bibliography{references}

\begin{thebibliography}{51}%
\makeatletter
\providecommand \@ifxundefined [1]{%
 \@ifx{#1\undefined}
}%
\providecommand \@ifnum [1]{%
 \ifnum #1\expandafter \@firstoftwo
 \else \expandafter \@secondoftwo
 \fi
}%
\providecommand \@ifx [1]{%
 \ifx #1\expandafter \@firstoftwo
 \else \expandafter \@secondoftwo
 \fi
}%
\providecommand \natexlab [1]{#1}%
\providecommand \enquote  [1]{``#1''}%
\providecommand \bibnamefont  [1]{#1}%
\providecommand \bibfnamefont [1]{#1}%
\providecommand \citenamefont [1]{#1}%
\providecommand \href@noop [0]{\@secondoftwo}%
\providecommand \href [0]{\begingroup \@sanitize@url \@href}%
\providecommand \@href[1]{\@@startlink{#1}\@@href}%
\providecommand \@@href[1]{\endgroup#1\@@endlink}%
\providecommand \@sanitize@url [0]{\catcode `\\12\catcode `\$12\catcode
  `\&12\catcode `\#12\catcode `\^12\catcode `\_12\catcode `\%12\relax}%
\providecommand \@@startlink[1]{}%
\providecommand \@@endlink[0]{}%
\providecommand \url  [0]{\begingroup\@sanitize@url \@url }%
\providecommand \@url [1]{\endgroup\@href {#1}{\urlprefix }}%
\providecommand \urlprefix  [0]{URL }%
\providecommand \Eprint [0]{\href }%
\providecommand \doibase [0]{http://dx.doi.org/}%
\providecommand \selectlanguage [0]{\@gobble}%
\providecommand \bibinfo  [0]{\@secondoftwo}%
\providecommand \bibfield  [0]{\@secondoftwo}%
\providecommand \translation [1]{[#1]}%
\providecommand \BibitemOpen [0]{}%
\providecommand \bibitemStop [0]{}%
\providecommand \bibitemNoStop [0]{.\EOS\space}%
\providecommand \EOS [0]{\spacefactor3000\relax}%
\providecommand \BibitemShut  [1]{\csname bibitem#1\endcsname}%
\let\auto@bib@innerbib\@empty
%</preamble>
\bibitem [{\citenamefont {Brandan}\ and\ \citenamefont
  {Satchler}(1997)}]{Bran97heavyion}%
  \BibitemOpen
  \bibfield  {author} {\bibinfo {author} {\bibfnamefont {M.}~\bibnamefont
  {Brandan}}\ and\ \bibinfo {author} {\bibfnamefont {G.}~\bibnamefont
  {Satchler}},\ }\href@noop {} {\bibfield  {journal} {\bibinfo  {journal}
  {Phys. Rep.}\ }\textbf {\bibinfo {volume} {285}},\ \bibinfo {pages} {143}
  (\bibinfo {year} {1997})}\BibitemShut {NoStop}%
\bibitem [{\citenamefont {Vorabbi}\ \emph {et~al.}(2018)\citenamefont
  {Vorabbi}, \citenamefont {Finelli},\ and\ \citenamefont
  {Giusti}}]{Vora18OpPot}%
  \BibitemOpen
  \bibfield  {author} {\bibinfo {author} {\bibfnamefont {M.}~\bibnamefont
  {Vorabbi}}, \bibinfo {author} {\bibfnamefont {P.}~\bibnamefont {Finelli}}, \
  and\ \bibinfo {author} {\bibfnamefont {C.}~\bibnamefont {Giusti}},\
  }\href@noop {} {\bibfield  {journal} {\bibinfo  {journal} {Phys. Rev. C}\
  }\textbf {\bibinfo {volume} {98}},\ \bibinfo {pages} {064602} (\bibinfo
  {year} {2018})}\BibitemShut {NoStop}%
\bibitem [{\citenamefont {Idini}\ \emph {et~al.}(2019)\citenamefont {Idini},
  \citenamefont {Barbieri},\ and\ \citenamefont {Navr\'atil}}]{Idin19OpPot}%
  \BibitemOpen
  \bibfield  {author} {\bibinfo {author} {\bibfnamefont {A.}~\bibnamefont
  {Idini}}, \bibinfo {author} {\bibfnamefont {C.}~\bibnamefont {Barbieri}}, \
  and\ \bibinfo {author} {\bibfnamefont {P.}~\bibnamefont {Navr\'atil}},\
  }\href@noop {} {\bibfield  {journal} {\bibinfo  {journal} {Phys. Rev. Lett.}\
  }\textbf {\bibinfo {volume} {123}},\ \bibinfo {pages} {092501} (\bibinfo
  {year} {2019})}\BibitemShut {NoStop}%
\bibitem [{\citenamefont {Rotureau}\ \emph {et~al.}(2020)\citenamefont
  {Rotureau}, \citenamefont {Danielewicz}, \citenamefont {Hagen}, \citenamefont
  {Jansen}, \citenamefont {Nunes},\ and\ \citenamefont
  {Papenbrock}}]{Rotu20OpPot}%
  \BibitemOpen
  \bibfield  {author} {\bibinfo {author} {\bibfnamefont {J.}~\bibnamefont
  {Rotureau}}, \bibinfo {author} {\bibfnamefont {P.}~\bibnamefont
  {Danielewicz}}, \bibinfo {author} {\bibfnamefont {G.}~\bibnamefont {Hagen}},
  \bibinfo {author} {\bibfnamefont {G.}~\bibnamefont {Jansen}}, \bibinfo
  {author} {\bibfnamefont {F.}~\bibnamefont {Nunes}}, \ and\ \bibinfo {author}
  {\bibfnamefont {T.}~\bibnamefont {Papenbrock}},\ }\href@noop {} {\bibfield
  {journal} {\bibinfo  {journal} {Springer Proc. Phys.}\ }\textbf {\bibinfo
  {volume} {238}},\ \bibinfo {pages} {183} (\bibinfo {year}
  {2020})}\BibitemShut {NoStop}%
\bibitem [{\citenamefont {Whitehead}\ \emph {et~al.}(2021)\citenamefont
  {Whitehead}, \citenamefont {Lim},\ and\ \citenamefont {Holt}}]{White20OpPot}%
  \BibitemOpen
  \bibfield  {author} {\bibinfo {author} {\bibfnamefont {T.~R.}\ \bibnamefont
  {Whitehead}}, \bibinfo {author} {\bibfnamefont {Y.}~\bibnamefont {Lim}}, \
  and\ \bibinfo {author} {\bibfnamefont {J.~W.}\ \bibnamefont {Holt}},\
  }\href@noop {} {\bibfield  {journal} {\bibinfo  {journal} {Phys. Rev. Lett.}\
  }\textbf {\bibinfo {volume} {127}},\ \bibinfo {pages} {182502} (\bibinfo
  {year} {2021})}\BibitemShut {NoStop}%
\bibitem [{\citenamefont {Epelbaum}\ \emph {et~al.}(2009)\citenamefont
  {Epelbaum}, \citenamefont {Hammer},\ and\ \citenamefont
  {Mei{\ss}ner}}]{Epel09RMP}%
  \BibitemOpen
  \bibfield  {author} {\bibinfo {author} {\bibfnamefont {E.}~\bibnamefont
  {Epelbaum}}, \bibinfo {author} {\bibfnamefont {H.-W.}\ \bibnamefont
  {Hammer}}, \ and\ \bibinfo {author} {\bibfnamefont {U.~G.}\ \bibnamefont
  {Mei{\ss}ner}},\ }\href@noop {} {\bibfield  {journal} {\bibinfo  {journal}
  {Rev. Mod. Phys.}\ }\textbf {\bibinfo {volume} {81}},\ \bibinfo {pages}
  {1773} (\bibinfo {year} {2009})}\BibitemShut {NoStop}%
%%CITATION = ARXIV:0811.1338;%%
\bibitem [{\citenamefont {Machleidt}\ and\ \citenamefont
  {Entem}(2011)}]{Mach11PR}%
  \BibitemOpen
  \bibfield  {author} {\bibinfo {author} {\bibfnamefont {R.}~\bibnamefont
  {Machleidt}}\ and\ \bibinfo {author} {\bibfnamefont {D.~R.}\ \bibnamefont
  {Entem}},\ }\href@noop {} {\bibfield  {journal} {\bibinfo  {journal} {Phys.
  Rep.}\ }\textbf {\bibinfo {volume} {503}},\ \bibinfo {pages} {1} (\bibinfo
  {year} {2011})}\BibitemShut {NoStop}%
%%CITATION = ARXIV:1105.2919;%%
\bibitem [{\citenamefont {Hammer}\ \emph {et~al.}(2013)\citenamefont {Hammer},
  \citenamefont {Nogga},\ and\ \citenamefont {Schwenk}}]{Hamm13RMP}%
  \BibitemOpen
  \bibfield  {author} {\bibinfo {author} {\bibfnamefont {H.-W.}\ \bibnamefont
  {Hammer}}, \bibinfo {author} {\bibfnamefont {A.}~\bibnamefont {Nogga}}, \
  and\ \bibinfo {author} {\bibfnamefont {A.}~\bibnamefont {Schwenk}},\
  }\href@noop {} {\bibfield  {journal} {\bibinfo  {journal} {Rev. Mod. Phys.}\
  }\textbf {\bibinfo {volume} {85}},\ \bibinfo {pages} {197} (\bibinfo {year}
  {2013})}\BibitemShut {NoStop}%
%%CITATION = ARXIV:1210.4273;%%
\bibitem [{\citenamefont {Chamon}\ \emph {et~al.}(2002)\citenamefont {Chamon},
  \citenamefont {Carlson}, \citenamefont {Gasques}, \citenamefont {Pereira},
  \citenamefont {De~Conti}, \citenamefont {Alvarez}, \citenamefont {Hussein},
  \citenamefont {C{\^a}ndido~Ribeiro}, \citenamefont {Rossi},\ and\
  \citenamefont {Silva}}]{Cham02SPdens}%
  \BibitemOpen
  \bibfield  {author} {\bibinfo {author} {\bibfnamefont {L.~C.}\ \bibnamefont
  {Chamon}}, \bibinfo {author} {\bibfnamefont {B.~V.}\ \bibnamefont {Carlson}},
  \bibinfo {author} {\bibfnamefont {L.~R.}\ \bibnamefont {Gasques}}, \bibinfo
  {author} {\bibfnamefont {D.}~\bibnamefont {Pereira}}, \bibinfo {author}
  {\bibfnamefont {C.}~\bibnamefont {De~Conti}}, \bibinfo {author}
  {\bibfnamefont {M.~A.~G.}\ \bibnamefont {Alvarez}}, \bibinfo {author}
  {\bibfnamefont {M.~S.}\ \bibnamefont {Hussein}}, \bibinfo {author}
  {\bibfnamefont {M.~A.}\ \bibnamefont {C{\^a}ndido~Ribeiro}}, \bibinfo
  {author} {\bibfnamefont {E.~S.}\ \bibnamefont {Rossi}, \bibfnamefont {Jr.}},
  \ and\ \bibinfo {author} {\bibfnamefont {C.~P.}\ \bibnamefont {Silva}},\
  }\href@noop {} {\bibfield  {journal} {\bibinfo  {journal} {Phys. Rev. C}\
  }\textbf {\bibinfo {volume} {66}},\ \bibinfo {pages} {014610} (\bibinfo
  {year} {2002})}\BibitemShut {NoStop}%
%%CITATION = NUCL-TH/0202015;%%
\bibitem [{\citenamefont {Pereira}\ \emph {et~al.}(2009)\citenamefont
  {Pereira}, \citenamefont {Lubian}, \citenamefont {Oliveira}, \citenamefont
  {de~Sousa},\ and\ \citenamefont {Chamon}}]{Pere09ImDFPot}%
  \BibitemOpen
  \bibfield  {author} {\bibinfo {author} {\bibfnamefont {D.}~\bibnamefont
  {Pereira}}, \bibinfo {author} {\bibfnamefont {J.}~\bibnamefont {Lubian}},
  \bibinfo {author} {\bibfnamefont {J.~R.~B.}\ \bibnamefont {Oliveira}},
  \bibinfo {author} {\bibfnamefont {D.~P.}\ \bibnamefont {de~Sousa}}, \ and\
  \bibinfo {author} {\bibfnamefont {L.~C.}\ \bibnamefont {Chamon}},\
  }\href@noop {} {\bibfield  {journal} {\bibinfo  {journal} {Phys. Lett. B}\
  }\textbf {\bibinfo {volume} {670}},\ \bibinfo {pages} {330} (\bibinfo {year}
  {2009})}\BibitemShut {NoStop}%
%%CITATION = PHLTA,B670,330;%%
\bibitem [{\citenamefont {Furumoto}\ \emph {et~al.}(2012)\citenamefont
  {Furumoto}, \citenamefont {Horiuchi}, \citenamefont {Takashina},
  \citenamefont {Yamamoto},\ and\ \citenamefont {Sakuragi}}]{Furu12OpPot}%
  \BibitemOpen
  \bibfield  {author} {\bibinfo {author} {\bibfnamefont {T.}~\bibnamefont
  {Furumoto}}, \bibinfo {author} {\bibfnamefont {W.}~\bibnamefont {Horiuchi}},
  \bibinfo {author} {\bibfnamefont {M.}~\bibnamefont {Takashina}}, \bibinfo
  {author} {\bibfnamefont {Y.}~\bibnamefont {Yamamoto}}, \ and\ \bibinfo
  {author} {\bibfnamefont {Y.}~\bibnamefont {Sakuragi}},\ }\href@noop {}
  {\bibfield  {journal} {\bibinfo  {journal} {Phys. Rev. C}\ }\textbf {\bibinfo
  {volume} {85}},\ \bibinfo {pages} {044607} (\bibinfo {year}
  {2012})}\BibitemShut {NoStop}%
%%CITATION = ARXIV:1204.2301;%%
\bibitem [{\citenamefont {Minomo}\ \emph {et~al.}(2016)\citenamefont {Minomo},
  \citenamefont {Kohno},\ and\ \citenamefont {Ogata}}]{Mino15ChEFTCC}%
  \BibitemOpen
  \bibfield  {author} {\bibinfo {author} {\bibfnamefont {K.}~\bibnamefont
  {Minomo}}, \bibinfo {author} {\bibfnamefont {M.}~\bibnamefont {Kohno}}, \
  and\ \bibinfo {author} {\bibfnamefont {K.}~\bibnamefont {Ogata}},\
  }\href@noop {} {\bibfield  {journal} {\bibinfo  {journal} {Phys. Rev. C}\
  }\textbf {\bibinfo {volume} {93}},\ \bibinfo {pages} {014607} (\bibinfo
  {year} {2016})}\BibitemShut {NoStop}%
%%CITATION = ARXIV:1509.04459;%%
\bibitem [{\citenamefont {Khoa}\ \emph {et~al.}(2016)\citenamefont {Khoa},
  \citenamefont {Phuc}, \citenamefont {Loan},\ and\ \citenamefont
  {Loc}}]{Khoa16DFM}%
  \BibitemOpen
  \bibfield  {author} {\bibinfo {author} {\bibfnamefont {D.~T.}\ \bibnamefont
  {Khoa}}, \bibinfo {author} {\bibfnamefont {N.~H.}\ \bibnamefont {Phuc}},
  \bibinfo {author} {\bibfnamefont {D.~T.}\ \bibnamefont {Loan}}, \ and\
  \bibinfo {author} {\bibfnamefont {B.~M.}\ \bibnamefont {Loc}},\ }\href@noop
  {} {\bibfield  {journal} {\bibinfo  {journal} {Phys. Rev. C}\ }\textbf
  {\bibinfo {volume} {94}},\ \bibinfo {pages} {034612} (\bibinfo {year}
  {2016})}\BibitemShut {NoStop}%
\bibitem [{\citenamefont {Durant}\ \emph {et~al.}(2018)\citenamefont {Durant},
  \citenamefont {Capel}, \citenamefont {Huth}, \citenamefont {Balantekin},\
  and\ \citenamefont {Schwenk}}]{Dura17DFP}%
  \BibitemOpen
  \bibfield  {author} {\bibinfo {author} {\bibfnamefont {V.}~\bibnamefont
  {Durant}}, \bibinfo {author} {\bibfnamefont {P.}~\bibnamefont {Capel}},
  \bibinfo {author} {\bibfnamefont {L.}~\bibnamefont {Huth}}, \bibinfo {author}
  {\bibfnamefont {A.~B.}\ \bibnamefont {Balantekin}}, \ and\ \bibinfo {author}
  {\bibfnamefont {A.}~\bibnamefont {Schwenk}},\ }\href@noop {} {\bibfield
  {journal} {\bibinfo  {journal} {Phys. Lett. B}\ }\textbf {\bibinfo {volume}
  {782}},\ \bibinfo {pages} {668} (\bibinfo {year} {2018})}\BibitemShut
  {NoStop}%
%%CITATION = ARXIV:1708.02527;%%
\bibitem [{\citenamefont {Durant}\ \emph {et~al.}(2020)\citenamefont {Durant},
  \citenamefont {Capel},\ and\ \citenamefont {Schwenk}}]{Dura20ImDisp}%
  \BibitemOpen
  \bibfield  {author} {\bibinfo {author} {\bibfnamefont {V.}~\bibnamefont
  {Durant}}, \bibinfo {author} {\bibfnamefont {P.}~\bibnamefont {Capel}}, \
  and\ \bibinfo {author} {\bibfnamefont {A.}~\bibnamefont {Schwenk}},\
  }\href@noop {} {\bibfield  {journal} {\bibinfo  {journal} {Phys. Rev. C}\
  }\textbf {\bibinfo {volume} {102}},\ \bibinfo {pages} {014622} (\bibinfo
  {year} {2020})}\BibitemShut {NoStop}%
\bibitem [{\citenamefont {Gezerlis}\ \emph {et~al.}(2013)\citenamefont
  {Gezerlis}, \citenamefont {Tews}, \citenamefont {Epelbaum}, \citenamefont
  {Gandolfi}, \citenamefont {Hebeler}, \citenamefont {Nogga},\ and\
  \citenamefont {Schwenk}}]{Geze13QMCchi}%
  \BibitemOpen
  \bibfield  {author} {\bibinfo {author} {\bibfnamefont {A.}~\bibnamefont
  {Gezerlis}}, \bibinfo {author} {\bibfnamefont {I.}~\bibnamefont {Tews}},
  \bibinfo {author} {\bibfnamefont {E.}~\bibnamefont {Epelbaum}}, \bibinfo
  {author} {\bibfnamefont {S.}~\bibnamefont {Gandolfi}}, \bibinfo {author}
  {\bibfnamefont {K.}~\bibnamefont {Hebeler}}, \bibinfo {author} {\bibfnamefont
  {A.}~\bibnamefont {Nogga}}, \ and\ \bibinfo {author} {\bibfnamefont
  {A.}~\bibnamefont {Schwenk}},\ }\href@noop {} {\bibfield  {journal} {\bibinfo
   {journal} {Phys. Rev. Lett.}\ }\textbf {\bibinfo {volume} {111}},\ \bibinfo
  {pages} {032501} (\bibinfo {year} {2013})}\BibitemShut {NoStop}%
%%CITATION = ARXIV:1303.6243;%%
\bibitem [{\citenamefont {Gezerlis}\ \emph {et~al.}(2014)\citenamefont
  {Gezerlis}, \citenamefont {Tews}, \citenamefont {Epelbaum}, \citenamefont
  {Freunek}, \citenamefont {Gandolfi}, \citenamefont {Hebeler}, \citenamefont
  {Nogga},\ and\ \citenamefont {Schwenk}}]{Geze14long}%
  \BibitemOpen
  \bibfield  {author} {\bibinfo {author} {\bibfnamefont {A.}~\bibnamefont
  {Gezerlis}}, \bibinfo {author} {\bibfnamefont {I.}~\bibnamefont {Tews}},
  \bibinfo {author} {\bibfnamefont {E.}~\bibnamefont {Epelbaum}}, \bibinfo
  {author} {\bibfnamefont {M.}~\bibnamefont {Freunek}}, \bibinfo {author}
  {\bibfnamefont {S.}~\bibnamefont {Gandolfi}}, \bibinfo {author}
  {\bibfnamefont {K.}~\bibnamefont {Hebeler}}, \bibinfo {author} {\bibfnamefont
  {A.}~\bibnamefont {Nogga}}, \ and\ \bibinfo {author} {\bibfnamefont
  {A.}~\bibnamefont {Schwenk}},\ }\href@noop {} {\bibfield  {journal} {\bibinfo
   {journal} {Phys. Rev. C}\ }\textbf {\bibinfo {volume} {90}},\ \bibinfo
  {pages} {054323} (\bibinfo {year} {2014})}\BibitemShut {NoStop}%
%%CITATION = ARXIV:1406.0454;%%
\bibitem [{\citenamefont {{H. De Vries and C. W. De Jager and C. De
  Vries}}(1987)}]{Devr87rhoel}%
  \BibitemOpen
  \bibfield  {author} {\bibinfo {author} {\bibnamefont {{H. De Vries and C. W.
  De Jager and C. De Vries}}},\ }\href@noop {} {\bibfield  {journal} {\bibinfo
  {journal} {At. Data Nucl. Data Tables}\ }\textbf {\bibinfo {volume} {36}},\
  \bibinfo {pages} {495} (\bibinfo {year} {1987})}\BibitemShut {NoStop}%
\bibitem [{\citenamefont {{M.A.G. Alvarez and L.C. Chamon and M.S. Hussein and
  D. Pereira and L.R. Gasques and E.S. Rossi Jr. and C.P.
  Silva}}(2003)}]{Alv03ImPot}%
  \BibitemOpen
  \bibfield  {author} {\bibinfo {author} {\bibnamefont {{M.A.G. Alvarez and
  L.C. Chamon and M.S. Hussein and D. Pereira and L.R. Gasques and E.S. Rossi
  Jr. and C.P. Silva}}},\ }\href@noop {} {\bibfield  {journal} {\bibinfo
  {journal} {Nuc. Phys. A}\ }\textbf {\bibinfo {volume} {723}},\ \bibinfo
  {pages} {93} (\bibinfo {year} {2003})}\BibitemShut {NoStop}%
\bibitem [{\citenamefont {Carlson}\ \emph {et~al.}(1990)\citenamefont
  {Carlson}, \citenamefont {Frederico}, \citenamefont {Hussein}, \citenamefont
  {Esbensen},\ and\ \citenamefont {Landowne}}]{Carl89Dispersion}%
  \BibitemOpen
  \bibfield  {author} {\bibinfo {author} {\bibfnamefont {B.~V.}\ \bibnamefont
  {Carlson}}, \bibinfo {author} {\bibfnamefont {T.}~\bibnamefont {Frederico}},
  \bibinfo {author} {\bibfnamefont {M.~S.}\ \bibnamefont {Hussein}}, \bibinfo
  {author} {\bibfnamefont {H.}~\bibnamefont {Esbensen}}, \ and\ \bibinfo
  {author} {\bibfnamefont {S.}~\bibnamefont {Landowne}},\ }\href@noop {}
  {\bibfield  {journal} {\bibinfo  {journal} {Phys. Rev. C}\ }\textbf {\bibinfo
  {volume} {41}},\ \bibinfo {pages} {933} (\bibinfo {year} {1990})}\BibitemShut
  {NoStop}%
\bibitem [{\citenamefont {Gonz{\'a}lez}\ and\ \citenamefont
  {Brandan}(2001)}]{Gonz01DisRel}%
  \BibitemOpen
  \bibfield  {author} {\bibinfo {author} {\bibfnamefont {M.~M.}\ \bibnamefont
  {Gonz{\'a}lez}}\ and\ \bibinfo {author} {\bibfnamefont {M.~E.}\ \bibnamefont
  {Brandan}},\ }\href@noop {} {\bibfield  {journal} {\bibinfo  {journal} {Nuc.
  Phys. A}\ }\textbf {\bibinfo {volume} {693}},\ \bibinfo {pages} {603}
  (\bibinfo {year} {2001})}\BibitemShut {NoStop}%
%%CITATION = NUPHA,A693,603;%%
\bibitem [{\citenamefont {{Ronald E. Brown and Y.C. Yang}}(1971)}]{Brow71aa}%
  \BibitemOpen
  \bibfield  {author} {\bibinfo {author} {\bibnamefont {{Ronald E. Brown and
  Y.C. Yang}}},\ }\href@noop {} {\bibfield  {journal} {\bibinfo  {journal}
  {Nuclear Physics A}\ }\textbf {\bibinfo {volume} {170}},\ \bibinfo {pages}
  {225} (\bibinfo {year} {1971})}\BibitemShut {NoStop}%
\bibitem [{\citenamefont {{ V.I. Kukulin and V.G. Neudatchin and Yu.F.
  Smirnov}}(1975)}]{Kuku75lightN}%
  \BibitemOpen
  \bibfield  {author} {\bibinfo {author} {\bibnamefont {{ V.I. Kukulin and V.G.
  Neudatchin and Yu.F. Smirnov}}},\ }\href@noop {} {\bibfield  {journal}
  {\bibinfo  {journal} {Nucl. Phys. A}\ }\textbf {\bibinfo {volume} {245}},\
  \bibinfo {pages} {429} (\bibinfo {year} {1975})}\BibitemShut {NoStop}%
\bibitem [{\citenamefont {{A.H. Al-Ghamdi and Awad A. Ibraheem, and M. El-Azab
  Farid}}(2015)}]{AlGha15aN}%
  \BibitemOpen
  \bibfield  {author} {\bibinfo {author} {\bibnamefont {{A.H. Al-Ghamdi and
  Awad A. Ibraheem, and M. El-Azab Farid}}},\ }\href@noop {} {\bibfield
  {journal} {\bibinfo  {journal} {Int. J. Mod. Phys. E}\ }\textbf {\bibinfo
  {volume} {24}},\ \bibinfo {pages} {1550003} (\bibinfo {year}
  {2015})}\BibitemShut {NoStop}%
\bibitem [{\citenamefont {Woo}\ \emph {et~al.}(1985)\citenamefont {Woo},
  \citenamefont {Kwiatkowski}, \citenamefont {Zhou},\ and\ \citenamefont
  {Viola}}]{Woo854He4He}%
  \BibitemOpen
  \bibfield  {author} {\bibinfo {author} {\bibfnamefont {L.~W.}\ \bibnamefont
  {Woo}}, \bibinfo {author} {\bibfnamefont {K.}~\bibnamefont {Kwiatkowski}},
  \bibinfo {author} {\bibfnamefont {S.~H.}\ \bibnamefont {Zhou}}, \ and\
  \bibinfo {author} {\bibfnamefont {V.~E.}\ \bibnamefont {Viola}},\ }\href@noop
  {} {\bibfield  {journal} {\bibinfo  {journal} {Phys. Rev. C}\ }\textbf
  {\bibinfo {volume} {32}},\ \bibinfo {pages} {706} (\bibinfo {year}
  {1985})}\BibitemShut {NoStop}%
\bibitem [{\citenamefont {Cowley}\ \emph {et~al.}(1994)\citenamefont {Cowley},
  \citenamefont {Steyn}, \citenamefont {F{\"o}rtsch}, \citenamefont {Lawrie},
  \citenamefont {Pilcher}, \citenamefont {Smit},\ and\ \citenamefont
  {Whittal}}]{Cow944He4He}%
  \BibitemOpen
  \bibfield  {author} {\bibinfo {author} {\bibfnamefont {A.~A.}\ \bibnamefont
  {Cowley}}, \bibinfo {author} {\bibfnamefont {G.~F.}\ \bibnamefont {Steyn}},
  \bibinfo {author} {\bibfnamefont {S.~V.}\ \bibnamefont {F{\"o}rtsch}},
  \bibinfo {author} {\bibfnamefont {J.~J.}\ \bibnamefont {Lawrie}}, \bibinfo
  {author} {\bibfnamefont {J.~V.}\ \bibnamefont {Pilcher}}, \bibinfo {author}
  {\bibfnamefont {F.~D.}\ \bibnamefont {Smit}}, \ and\ \bibinfo {author}
  {\bibfnamefont {D.~M.}\ \bibnamefont {Whittal}},\ }\href@noop {} {\bibfield
  {journal} {\bibinfo  {journal} {Phys. Rev. C}\ }\textbf {\bibinfo {volume}
  {50}},\ \bibinfo {pages} {2449} (\bibinfo {year} {1994})}\BibitemShut
  {NoStop}%
\bibitem [{\citenamefont {Rao}\ \emph {et~al.}(2000)\citenamefont {Rao},
  \citenamefont {Nadasen}, \citenamefont {Sisan}, \citenamefont {Yuhasz},
  \citenamefont {Mercer}, \citenamefont {Austin}, \citenamefont {Roos},\ and\
  \citenamefont {Warner}}]{Rao004He4He}%
  \BibitemOpen
  \bibfield  {author} {\bibinfo {author} {\bibfnamefont {K.~A.~G.}\
  \bibnamefont {Rao}}, \bibinfo {author} {\bibfnamefont {A.}~\bibnamefont
  {Nadasen}}, \bibinfo {author} {\bibfnamefont {D.}~\bibnamefont {Sisan}},
  \bibinfo {author} {\bibfnamefont {W.}~\bibnamefont {Yuhasz}}, \bibinfo
  {author} {\bibfnamefont {D.}~\bibnamefont {Mercer}}, \bibinfo {author}
  {\bibfnamefont {S.~M.}\ \bibnamefont {Austin}}, \bibinfo {author}
  {\bibfnamefont {P.~G.}\ \bibnamefont {Roos}}, \ and\ \bibinfo {author}
  {\bibfnamefont {R.~E.}\ \bibnamefont {Warner}},\ }\href@noop {} {\bibfield
  {journal} {\bibinfo  {journal} {Phys. Rev. C}\ }\textbf {\bibinfo {volume}
  {62}},\ \bibinfo {pages} {014607} (\bibinfo {year} {2000})}\BibitemShut
  {NoStop}%
\bibitem [{\citenamefont {Hauser}\ \emph {et~al.}(1969)\citenamefont {Hauser},
  \citenamefont {L\"ohken}, \citenamefont {Rebel}, \citenamefont {Schatz},
  \citenamefont {Schweimer},\ and\ \citenamefont {Specht}}]{Haus69Al104}%
  \BibitemOpen
  \bibfield  {author} {\bibinfo {author} {\bibfnamefont {G.}~\bibnamefont
  {Hauser}}, \bibinfo {author} {\bibfnamefont {R.}~\bibnamefont {L\"ohken}},
  \bibinfo {author} {\bibfnamefont {H.}~\bibnamefont {Rebel}}, \bibinfo
  {author} {\bibfnamefont {G.}~\bibnamefont {Schatz}}, \bibinfo {author}
  {\bibfnamefont {G.~W.}\ \bibnamefont {Schweimer}}, \ and\ \bibinfo {author}
  {\bibfnamefont {J.}~\bibnamefont {Specht}},\ }\href@noop {} {\bibfield
  {journal} {\bibinfo  {journal} {Nucl. Phys. A}\ }\textbf {\bibinfo {volume}
  {128}},\ \bibinfo {pages} {81} (\bibinfo {year} {1969})}\BibitemShut
  {NoStop}%
\bibitem [{\citenamefont {Gils}\ \emph {et~al.}(1980)\citenamefont {Gils},
  \citenamefont {Friedman}, \citenamefont {Rebel}, \citenamefont {Buschmann},
  \citenamefont {Zagromski}, \citenamefont {Klewe-Nebenius}, \citenamefont
  {Neumann}, \citenamefont {Pesl},\ and\ \citenamefont
  {Bechtold}}]{Gils804HeCa}%
  \BibitemOpen
  \bibfield  {author} {\bibinfo {author} {\bibfnamefont {H.~J.}\ \bibnamefont
  {Gils}}, \bibinfo {author} {\bibfnamefont {E.}~\bibnamefont {Friedman}},
  \bibinfo {author} {\bibfnamefont {H.}~\bibnamefont {Rebel}}, \bibinfo
  {author} {\bibfnamefont {J.}~\bibnamefont {Buschmann}}, \bibinfo {author}
  {\bibfnamefont {S.}~\bibnamefont {Zagromski}}, \bibinfo {author}
  {\bibfnamefont {H.}~\bibnamefont {Klewe-Nebenius}}, \bibinfo {author}
  {\bibfnamefont {B.}~\bibnamefont {Neumann}}, \bibinfo {author} {\bibfnamefont
  {R.}~\bibnamefont {Pesl}}, \ and\ \bibinfo {author} {\bibfnamefont
  {G.}~\bibnamefont {Bechtold}},\ }\href@noop {} {\bibfield  {journal}
  {\bibinfo  {journal} {Phys. Rev. C}\ }\textbf {\bibinfo {volume} {21}},\
  \bibinfo {pages} {1239} (\bibinfo {year} {1980})}\BibitemShut {NoStop}%
\bibitem [{\citenamefont {John}\ \emph {et~al.}(2003)\citenamefont {John},
  \citenamefont {Tokimoto}, \citenamefont {Lui}, \citenamefont {Clark},
  \citenamefont {Chen},\ and\ \citenamefont {Youngblood}}]{John034He12C}%
  \BibitemOpen
  \bibfield  {author} {\bibinfo {author} {\bibfnamefont {B.}~\bibnamefont
  {John}}, \bibinfo {author} {\bibfnamefont {Y.}~\bibnamefont {Tokimoto}},
  \bibinfo {author} {\bibfnamefont {Y.-W.}\ \bibnamefont {Lui}}, \bibinfo
  {author} {\bibfnamefont {H.~L.}\ \bibnamefont {Clark}}, \bibinfo {author}
  {\bibfnamefont {X.}~\bibnamefont {Chen}}, \ and\ \bibinfo {author}
  {\bibfnamefont {D.~H.}\ \bibnamefont {Youngblood}},\ }\href@noop {}
  {\bibfield  {journal} {\bibinfo  {journal} {Phys. Rev. C}\ }\textbf {\bibinfo
  {volume} {68}},\ \bibinfo {pages} {014305} (\bibinfo {year}
  {2003})}\BibitemShut {NoStop}%
\bibitem [{\citenamefont {Lui}\ \emph {et~al.}(2001)\citenamefont {Lui},
  \citenamefont {Clark},\ and\ \citenamefont {Youngblood}}]{Lui014He16O}%
  \BibitemOpen
  \bibfield  {author} {\bibinfo {author} {\bibfnamefont {Y.-W.}\ \bibnamefont
  {Lui}}, \bibinfo {author} {\bibfnamefont {H.~L.}\ \bibnamefont {Clark}}, \
  and\ \bibinfo {author} {\bibfnamefont {D.~H.}\ \bibnamefont {Youngblood}},\
  }\href@noop {} {\bibfield  {journal} {\bibinfo  {journal} {Phys. Rev. C}\
  }\textbf {\bibinfo {volume} {64}},\ \bibinfo {pages} {064308} (\bibinfo
  {year} {2001})}\BibitemShut {NoStop}%
\bibitem [{\citenamefont {Youngblood}\ \emph {et~al.}(1997)\citenamefont
  {Youngblood}, \citenamefont {Lui},\ and\ \citenamefont
  {Clark}}]{Youn974He40Ca}%
  \BibitemOpen
  \bibfield  {author} {\bibinfo {author} {\bibfnamefont {D.~H.}\ \bibnamefont
  {Youngblood}}, \bibinfo {author} {\bibfnamefont {Y.-W.}\ \bibnamefont {Lui}},
  \ and\ \bibinfo {author} {\bibfnamefont {H.~L.}\ \bibnamefont {Clark}},\
  }\href@noop {} {\bibfield  {journal} {\bibinfo  {journal} {Phys. Rev. C}\
  }\textbf {\bibinfo {volume} {55}},\ \bibinfo {pages} {2811} (\bibinfo {year}
  {1997})}\BibitemShut {NoStop}%
\bibitem [{\citenamefont {Lui}\ \emph {et~al.}(2011)\citenamefont {Lui},
  \citenamefont {Youngblood}, \citenamefont {Shlomo}, \citenamefont {Chen},
  \citenamefont {Tokimoto}, \citenamefont {Krishichayan}, \citenamefont
  {Anders},\ and\ \citenamefont {Button}}]{Lui114He48Ca}%
  \BibitemOpen
  \bibfield  {author} {\bibinfo {author} {\bibfnamefont {Y.-W.}\ \bibnamefont
  {Lui}}, \bibinfo {author} {\bibfnamefont {D.~H.}\ \bibnamefont {Youngblood}},
  \bibinfo {author} {\bibfnamefont {S.}~\bibnamefont {Shlomo}}, \bibinfo
  {author} {\bibfnamefont {X.}~\bibnamefont {Chen}}, \bibinfo {author}
  {\bibfnamefont {Y.}~\bibnamefont {Tokimoto}}, \bibinfo {author} {\bibnamefont
  {Krishichayan}}, \bibinfo {author} {\bibfnamefont {M.}~\bibnamefont
  {Anders}}, \ and\ \bibinfo {author} {\bibfnamefont {J.}~\bibnamefont
  {Button}},\ }\href@noop {} {\bibfield  {journal} {\bibinfo  {journal} {Phys.
  Rev. C}\ }\textbf {\bibinfo {volume} {83}},\ \bibinfo {pages} {044327}
  (\bibinfo {year} {2011})}\BibitemShut {NoStop}%
\bibitem [{\citenamefont {Li}\ \emph {et~al.}(2010)\citenamefont {Li},
  \citenamefont {Garg}, \citenamefont {Liu}, \citenamefont {Marks},
  \citenamefont {Nayak}, \citenamefont {Rao}, \citenamefont {Fujiwara},
  \citenamefont {Hashimoto}, \citenamefont {Nakanishi}, \citenamefont {Okumura}
  \emph {et~al.}}]{Li104He120Sn}%
  \BibitemOpen
  \bibfield  {author} {\bibinfo {author} {\bibfnamefont {T.}~\bibnamefont
  {Li}}, \bibinfo {author} {\bibfnamefont {U.}~\bibnamefont {Garg}}, \bibinfo
  {author} {\bibfnamefont {Y.}~\bibnamefont {Liu}}, \bibinfo {author}
  {\bibfnamefont {R.}~\bibnamefont {Marks}}, \bibinfo {author} {\bibfnamefont
  {B.}~\bibnamefont {Nayak}}, \bibinfo {author} {\bibfnamefont {P.~V.~M.}\
  \bibnamefont {Rao}}, \bibinfo {author} {\bibfnamefont {M.}~\bibnamefont
  {Fujiwara}}, \bibinfo {author} {\bibfnamefont {H.}~\bibnamefont {Hashimoto}},
  \bibinfo {author} {\bibfnamefont {K.}~\bibnamefont {Nakanishi}}, \bibinfo
  {author} {\bibfnamefont {S.}~\bibnamefont {Okumura}},  \emph {et~al.},\
  }\href@noop {} {\bibfield  {journal} {\bibinfo  {journal} {Phys. Rev. C}\
  }\textbf {\bibinfo {volume} {81}},\ \bibinfo {pages} {034309} (\bibinfo
  {year} {2010})}\BibitemShut {NoStop}%
\bibitem [{\citenamefont {Satchler}\ and\ \citenamefont
  {Love}(1979)}]{Satc79Folding}%
  \BibitemOpen
  \bibfield  {author} {\bibinfo {author} {\bibfnamefont {G.~R.}\ \bibnamefont
  {Satchler}}\ and\ \bibinfo {author} {\bibfnamefont {W.~G.}\ \bibnamefont
  {Love}},\ }\href@noop {} {\bibfield  {journal} {\bibinfo  {journal} {Phys.
  Rep.}\ }\textbf {\bibinfo {volume} {55}},\ \bibinfo {pages} {183 } (\bibinfo
  {year} {1979})}\BibitemShut {NoStop}%
\bibitem [{\citenamefont {Negele}\ and\ \citenamefont
  {Vautherin}(1972)}]{Nege72DME1}%
  \BibitemOpen
  \bibfield  {author} {\bibinfo {author} {\bibfnamefont {J.~W.}\ \bibnamefont
  {Negele}}\ and\ \bibinfo {author} {\bibfnamefont {D.}~\bibnamefont
  {Vautherin}},\ }\href@noop {} {\bibfield  {journal} {\bibinfo  {journal}
  {Phys. Rev. C}\ }\textbf {\bibinfo {volume} {5}},\ \bibinfo {pages} {1472}
  (\bibinfo {year} {1972})}\BibitemShut {NoStop}%
%%CITATION = PHRVA,C5,1472;%%
\bibitem [{\citenamefont {Bogner}\ \emph {et~al.}(2009)\citenamefont {Bogner},
  \citenamefont {Furnstahl},\ and\ \citenamefont {Platter}}]{Bogn09DME}%
  \BibitemOpen
  \bibfield  {author} {\bibinfo {author} {\bibfnamefont {S.~K.}\ \bibnamefont
  {Bogner}}, \bibinfo {author} {\bibfnamefont {R.~J.}\ \bibnamefont
  {Furnstahl}}, \ and\ \bibinfo {author} {\bibfnamefont {L.}~\bibnamefont
  {Platter}},\ }\href@noop {} {\bibfield  {journal} {\bibinfo  {journal} {Eur.
  Phys. J. A}\ }\textbf {\bibinfo {volume} {39}},\ \bibinfo {pages} {219}
  (\bibinfo {year} {2009})}\BibitemShut {NoStop}%
%%CITATION = ARXIV:0811.4198;%%
\bibitem [{\citenamefont {Lynn}\ \emph {et~al.}(2017)\citenamefont {Lynn},
  \citenamefont {Tews}, \citenamefont {Carlson}, \citenamefont {Gandolfi},
  \citenamefont {Gezerlis}, \citenamefont {Schmidt},\ and\ \citenamefont
  {Schwenk}}]{Lynn17QMClight}%
  \BibitemOpen
  \bibfield  {author} {\bibinfo {author} {\bibfnamefont {J.~E.}\ \bibnamefont
  {Lynn}}, \bibinfo {author} {\bibfnamefont {I.}~\bibnamefont {Tews}}, \bibinfo
  {author} {\bibfnamefont {J.}~\bibnamefont {Carlson}}, \bibinfo {author}
  {\bibfnamefont {S.}~\bibnamefont {Gandolfi}}, \bibinfo {author}
  {\bibfnamefont {A.}~\bibnamefont {Gezerlis}}, \bibinfo {author}
  {\bibfnamefont {K.~E.}\ \bibnamefont {Schmidt}}, \ and\ \bibinfo {author}
  {\bibfnamefont {A.}~\bibnamefont {Schwenk}},\ }\href@noop {} {\bibfield
  {journal} {\bibinfo  {journal} {Phys. Rev. C}\ }\textbf {\bibinfo {volume}
  {96}},\ \bibinfo {pages} {054007} (\bibinfo {year} {2017})}\BibitemShut
  {NoStop}%
%%CITATION = ARXIV:1706.07668;%%
\bibitem [{\citenamefont {Chen}\ and\ \citenamefont
  {Piekarewicz}(2015)}]{Chen15NStars}%
  \BibitemOpen
  \bibfield  {author} {\bibinfo {author} {\bibfnamefont {W.-C.}\ \bibnamefont
  {Chen}}\ and\ \bibinfo {author} {\bibfnamefont {J.}~\bibnamefont
  {Piekarewicz}},\ }\href@noop {} {\bibfield  {journal} {\bibinfo  {journal}
  {Phys. Rev. Lett.}\ }\textbf {\bibinfo {volume} {115}},\ \bibinfo {pages}
  {161101} (\bibinfo {year} {2015})}\BibitemShut {NoStop}%
\bibitem [{\citenamefont {Feshbach}(1958)}]{Fesh58}%
  \BibitemOpen
  \bibfield  {author} {\bibinfo {author} {\bibfnamefont {H.}~\bibnamefont
  {Feshbach}},\ }\href@noop {} {\bibfield  {journal} {\bibinfo  {journal} {Ann.
  Phys.}\ }\textbf {\bibinfo {volume} {5}},\ \bibinfo {pages} {357} (\bibinfo
  {year} {1958})}\BibitemShut {NoStop}%
\bibitem [{\citenamefont {de~L.~Kronig}(1926)}]{Kron26Disp}%
  \BibitemOpen
  \bibfield  {author} {\bibinfo {author} {\bibfnamefont {R.}~\bibnamefont
  {de~L.~Kronig}},\ }\href@noop {} {\bibfield  {journal} {\bibinfo  {journal}
  {J. Opt. Soc. Am.}\ }\textbf {\bibinfo {volume} {12}},\ \bibinfo {pages}
  {547} (\bibinfo {year} {1926})}\BibitemShut {NoStop}%
\bibitem [{\citenamefont {Kramers}(1927)}]{Kram27Disp}%
  \BibitemOpen
  \bibfield  {author} {\bibinfo {author} {\bibfnamefont {H.~A.}\ \bibnamefont
  {Kramers}},\ }\href@noop {} {\bibfield  {journal} {\bibinfo  {journal} {Atti
  Cong. Intern. Fisica (Transactions of Volta Centenary Congress)}\ }\textbf
  {\bibinfo {volume} {2}},\ \bibinfo {pages} {545} (\bibinfo {year}
  {1927})}\BibitemShut {NoStop}%
\bibitem [{\citenamefont {Feshbach}(1962)}]{Fesh62}%
  \BibitemOpen
  \bibfield  {author} {\bibinfo {author} {\bibfnamefont {H.}~\bibnamefont
  {Feshbach}},\ }\href@noop {} {\bibfield  {journal} {\bibinfo  {journal} {Ann.
  Phys.}\ }\textbf {\bibinfo {volume} {19}},\ \bibinfo {pages} {287} (\bibinfo
  {year} {1962})}\BibitemShut {NoStop}%
\bibitem [{\citenamefont {Passatore}(1967)}]{Pass67DispRel}%
  \BibitemOpen
  \bibfield  {author} {\bibinfo {author} {\bibfnamefont {G.}~\bibnamefont
  {Passatore}},\ }\href@noop {} {\bibfield  {journal} {\bibinfo  {journal}
  {Nuc. Phys. A}\ }\textbf {\bibinfo {volume} {95}},\ \bibinfo {pages} {694}
  (\bibinfo {year} {1967})}\BibitemShut {NoStop}%
\bibitem [{\citenamefont {Mahaux}\ \emph {et~al.}(1986)\citenamefont {Mahaux},
  \citenamefont {Ng{\^o}},\ and\ \citenamefont {Satchler}}]{Maha86DispRel}%
  \BibitemOpen
  \bibfield  {author} {\bibinfo {author} {\bibfnamefont {C.}~\bibnamefont
  {Mahaux}}, \bibinfo {author} {\bibfnamefont {H.}~\bibnamefont {Ng{\^o}}}, \
  and\ \bibinfo {author} {\bibfnamefont {G.~R.}\ \bibnamefont {Satchler}},\
  }\href@noop {} {\bibfield  {journal} {\bibinfo  {journal} {Nucl. Phys. A}\
  }\textbf {\bibinfo {volume} {449}},\ \bibinfo {pages} {354} (\bibinfo {year}
  {1986})}\BibitemShut {NoStop}%
\bibitem [{\citenamefont {Horsley}\ \emph {et~al.}(2015)\citenamefont
  {Horsley}, \citenamefont {Artoni},\ and\ \citenamefont
  {La~Rocca}}]{Hors15SpatKramKro}%
  \BibitemOpen
  \bibfield  {author} {\bibinfo {author} {\bibfnamefont {S.~A.~R.}\
  \bibnamefont {Horsley}}, \bibinfo {author} {\bibfnamefont {M.}~\bibnamefont
  {Artoni}}, \ and\ \bibinfo {author} {\bibfnamefont {G.~C.}\ \bibnamefont
  {La~Rocca}},\ }\href@noop {} {\bibfield  {journal} {\bibinfo  {journal}
  {Nature Photonics}\ }\textbf {\bibinfo {volume} {9}},\ \bibinfo {pages} {436}
  (\bibinfo {year} {2015})}\BibitemShut {NoStop}%
\bibitem [{\citenamefont {Mohr}\ \emph {et~al.}(2016)\citenamefont {Mohr},
  \citenamefont {Newell},\ and\ \citenamefont {Taylor}}]{Mohr15CODATA}%
  \BibitemOpen
  \bibfield  {author} {\bibinfo {author} {\bibfnamefont {P.~J.}\ \bibnamefont
  {Mohr}}, \bibinfo {author} {\bibfnamefont {D.~B.}\ \bibnamefont {Newell}}, \
  and\ \bibinfo {author} {\bibfnamefont {B.~N.}\ \bibnamefont {Taylor}},\
  }\href@noop {} {\bibfield  {journal} {\bibinfo  {journal} {Rev. Mod. Phys.}\
  }\textbf {\bibinfo {volume} {88}},\ \bibinfo {pages} {035009} (\bibinfo
  {year} {2016})}\BibitemShut {NoStop}%
\bibitem [{\citenamefont {Papoulias}\ and\ \citenamefont
  {Kosmas}(2015)}]{Papo15neutr}%
  \BibitemOpen
  \bibfield  {author} {\bibinfo {author} {\bibfnamefont {D.~K.}\ \bibnamefont
  {Papoulias}}\ and\ \bibinfo {author} {\bibfnamefont {T.~S.}\ \bibnamefont
  {Kosmas}},\ }\href@noop {} {\bibfield  {journal} {\bibinfo  {journal} {Adv.
  High Energy Phys.}\ }\textbf {\bibinfo {volume} {2015}},\ \bibinfo {pages}
  {763648} (\bibinfo {year} {2015})}\BibitemShut {NoStop}%
\bibitem [{\citenamefont {{G. Hagen, A. Ekstr{\"o}m, C. Forss{\'e}n, G. R.
  Jansen, W. Nazarewicz, T. Papenbrock, K. A. Wendt, S. Bacca, N. Barnea, B.
  Carlsson, C. Drischler, K. Hebeler, M. Hjorth-Jensen, M. Miorelli, G.
  Orlandini, A. Schwenk and J. Simonis}}(2016)}]{Hage16NatPhys}%
  \BibitemOpen
  \bibfield  {author} {\bibinfo {author} {\bibnamefont {{G. Hagen, A.
  Ekstr{\"o}m, C. Forss{\'e}n, G. R. Jansen, W. Nazarewicz, T. Papenbrock, K.
  A. Wendt, S. Bacca, N. Barnea, B. Carlsson, C. Drischler, K. Hebeler, M.
  Hjorth-Jensen, M. Miorelli, G. Orlandini, A. Schwenk and J. Simonis}}},\
  }\href@noop {} {\bibfield  {journal} {\bibinfo  {journal} {Nature Phys.}\
  }\textbf {\bibinfo {volume} {12}},\ \bibinfo {pages} {186} (\bibinfo {year}
  {2016})}\BibitemShut {NoStop}%
%%CITATION = ARXIV:1509.07169;%%
\bibitem [{\citenamefont {Nolte}\ \emph {et~al.}(1987)\citenamefont {Nolte},
  \citenamefont {Machner},\ and\ \citenamefont {Bojowald}}]{Nolt87GOP}%
  \BibitemOpen
  \bibfield  {author} {\bibinfo {author} {\bibfnamefont {M.}~\bibnamefont
  {Nolte}}, \bibinfo {author} {\bibfnamefont {H.}~\bibnamefont {Machner}}, \
  and\ \bibinfo {author} {\bibfnamefont {J.}~\bibnamefont {Bojowald}},\
  }\href@noop {} {\bibfield  {journal} {\bibinfo  {journal} {Phys. Rev. C}\
  }\textbf {\bibinfo {volume} {36}},\ \bibinfo {pages} {1312} (\bibinfo {year}
  {1987})}\BibitemShut {NoStop}%
\bibitem [{\citenamefont {Weber}(2018)}]{Webe17TFP}%
  \BibitemOpen
  \bibfield  {author} {\bibinfo {author} {\bibfnamefont {S.}~\bibnamefont
  {Weber}},\ }\emph {\bibinfo {title} {{Three-body forces and nucleus-nucleus
  interactions}}},\ \href@noop {} {\bibinfo {type} {{B.Sc. Thesis}}},\ \bibinfo
   {school} {{Technische Universit{\"a}t Darmstadt}} (\bibinfo {year}
  {2018})\BibitemShut {NoStop}%
\end{thebibliography}%

\end{document}